\documentclass[11pt,english,american]{article}
\usepackage{helvet}

\usepackage[T1]{fontenc}
\usepackage[latin1]{inputenc}
\usepackage[a4paper]{geometry}
\usepackage{color}
\usepackage{units}
\usepackage{amsmath}
\usepackage{amssymb}
\usepackage{graphicx}
\usepackage{setspace}
\usepackage[numbers]{natbib}
\makeatletter

\providecommand{\tabularnewline}{\\}

\usepackage{mdwlist}
\usepackage{amsmath}

\usepackage{colortbl}
\usepackage{changepage}
\usepackage{textcomp,marvosym}
\usepackage[font={footnotesize,it}]{caption}

\makeatletter
\def\blfootnote{\gdef\@thefnmark{}\@footnotetext}
\makeatother

\usepackage[a4paper]{geometry}

\usepackage[american]{babel}

\date{}

\makeatother

\usepackage{babel}
\begin{document}
\global\long\def\G{\Gamma}
\global\long\def\t{\theta}
\global\long\def\s{\text{s}}

 \begin{flushleft}
{\Large
\textbf\newline{Learning universal computations with spikes}
}
\newline
\\
Dominik Thalmeier$^{1}$, Marvin Uhlmann$^{2,3}$, Hilbert J. Kappen$^{1}$,\\
Raoul-Martin Memmesheimer$^{4,3,*}$

\
\bf{1} Donders Institute, Department of Biophysics, Radboud University,
Nijmegen, Netherlands.

\bf{2} Max Planck Institute for Psycholinguistics, Department for
Neurobiology of Language, Nijmegen, Netherlands.

\bf{3} Donders Institute, Department for Neuroinformatics, Radboud
University, Nijmegen, Netherlands.

\bf{4} Center for Theoretical Neuroscience, Columbia University, New
York, USA.
* E-mail: rm3354@cumc.columbia.edu
\end{flushleft}

\section*{Abstract}
Providing the neurobiological basis of information processing in higher
animals, spiking neural networks must be able to learn a variety of
complicated computations, including the generation of appropriate,
possibly delayed reactions to inputs and the self-sustained generation
of complex activity patterns, e.g.~for locomotion. Many such computations
require previous building of intrinsic world models. Here we show
how spiking neural networks may solve these different tasks. Firstly,
we derive constraints under which classes of spiking neural networks
lend themselves to substrates of powerful general purpose computing.
The networks contain dendritic or synaptic nonlinearities and have
a constrained connectivity. We then combine such networks with learning
rules for outputs or recurrent connections. We show that this allows
to learn even difficult benchmark tasks such as the self-sustained
generation of desired low-dimensional chaotic dynamics or memory-dependent
computations. Furthermore, we show how spiking networks can build
models of external world systems and use the acquired knowledge to
control them.

\section*{Author Summary}
Animals and humans can learn versatile computations such as the generation
of complicated activity patterns to steer movements or the generation
of appropriate outputs in response to inputs. Such learning must be
accomplished by networks of nerve cells in the brain, which communicate
with short electrical impulses, so-called spikes. Here we show how
such networks may perform the learning. We track their ability back
to experimentally found nonlinearities in the couplings between nerve
cells and to a network connectivity that complies with constraints.
We show that the spiking networks are able to learn difficult tasks
such as the generation of desired chaotic activity and the prediction
of the impact of actions on the environment. The latter allows to
compute optimal actions by mental exploration.

\section*{Introduction}

The understanding of neural network dynamics on the mesoscopic level
of hundreds and thousands of neurons and their ability to learn highly
complicated computations is a fundamental open challenge in neuroscience.
For biological systems, such an understanding will allow to connect
the microscopic level of single neurons and the macroscopic level
of cognition and behavior. In artificial computing, it may allow to
propose new, possibly more efficient computing schemes.

Randomly connected mesoscopic networks can be a suitable substrate
for computations \cite{MNM02,JH04,maass2007computational,SA09,sussillo2013opening},
as they reflect the input in a complicated, nonlinear way and at the
same time maintain, like a computational ``reservoir'', fading memory
of past inputs as well as of transformations and combinations of them.
This includes the results of computations on current and past inputs.
Simple readout neurons may then learn to extract the desired result;
the computations are executed in real time, i.e.~without the need
to wait for convergence to an attractor (``reservoir computing'')
\cite{MNM02,JH04}. Non-random and adaptive network connectivity can
change performance \cite{jaeger2009reservoir,Lazar2009,klampfl2013emergence}.

Networks with higher computational power, in particular with the additional
ability to learn self-sustained patterns of activity and persistent
memory, require an output feedback or equivalent learning of their
recurrent connections \cite{JH04,maass2007computational}. However,
network modeling approaches achieving such universal (i.e.~general
purpose) computational capabilities so far concentrated on networks
of continuous rate units \cite{JH04,SA09}, which do not take into
account the characteristics that neurons in biological neural networks
communicate via spikes. Indeed, the dynamics of spiking neural networks
are discontinuous, usually highly chaotic, variable, and noisy. Readouts
of such spiking networks show low signal-to-noise ratios. This hinders
computations following the described principle in particular in presence
of feedback or equivalent plastic recurrent connections, and has questioned
it as model for computations in biological neural systems \cite{joshi2005movement,mayor2005signal,wallace2013randomly}.

Here we first introduce a class of recurrent spiking neural networks
that are suited as a substrate to learn universal computations. They
are based on standard, established neuron models, take into account
synaptic or dendritic nonlinearities and are required to respect some
structural constraints regarding the connectivity of the network.
To derive them we employ a precise spike coding scheme similar to
ref.~\cite{boerlin2013predictive}, which was introduced to approximate
linear continuous dynamics.

Thereafter we endow the introduced spiking networks with learning
rules for either the output or the recurrent connection weights and
show that this enables them to learn equally complicated, memory dependent
computations as non-spiking continuous rate networks. The spiking
networks we are using have only medium sizes, between tens and a few
thousands of neurons, like networks of rate neurons employed for similar
tasks. We demonstrate the capabilities of our networks by applying
them to challenging learning problems which are of importance in biological
contexts. In particular, we show how spiking neural networks can learn
the self-sustained generation of complicated dynamical patterns, and
how they can build world models, which allow to compute optimal actions
to appropriately influence an environment.

\section*{Results}

\subsection*{Continuous signal coding spiking neural networks (CSNs)}

\subsubsection*{Network architecture}

For our study, we use leaky integrate-and-fire neurons. These incorporate
crucial features of biological neurons, such as operation in continuous
time, spike generation and reset, while also maintaining some degree
of analytical tractability. A network consists of $N$ neurons. The
state of a neuron $n$ is given by its membrane potential $V_{n}(t)$.
The membrane potential performs a leaky integration of the input and
a spike is generated when $V_{n}(t)$ reaches a threshold, resulting
in a spiketrain 
\begin{equation}
s_{n}(t)=\sum_{t_{n}}\delta(t-t_{n})\label{eq:st}
\end{equation}
with spike times $t_{n}$ and the Dirac delta-distribution $\delta$.
After a spike, the neuron is reset to the reset potential, which lies
$\theta$ below the threshold. The spike train generates a train of
exponentially decaying normalized synaptic currents 
\begin{equation}
r_{n}(t)=\sum_{t_{n}}e^{-\lambda_{s}(t-t_{n})}\Theta(t-t_{n})\,\,\Leftrightarrow\,\,\dot{r}_{n}(t)=-\lambda_{s}r_{n}(t)+s_{n}(t),\label{eq:rt}
\end{equation}
where $\tau_{s}=\lambda_{s}^{-1}$ is the time constant of the synaptic
decay and $\Theta(\,.\,)$ is the Heaviside theta-function.

Throughout the article we consider two closely related types of neurons,
neurons with saturating synapses and neurons with nonlinear dendrites
(cf.~Fig.~1). In the model with saturating synapses (Fig.~1a),
the membrane potential $V_{n}(t)$ of neuron $n$ obeys 
\begin{align}
\dot{V}_{n}(t)= & -\lambda_{V}V_{n}(t)+\sum_{m=1}^{N}A_{nm}\tanh\left(\gamma r_{m}(t)\right)+V_{\text{r}}\lambda_{s}r_{n}(t)\nonumber \\
 & -\t s_{n}(t)+I_{\text{e,}n}(t),\label{eq:Vdot}
\end{align}
with membrane time constant $\tau_{m}=\lambda_{V}^{-1}$.

\begin{figure}
\includegraphics[width=1\textwidth]{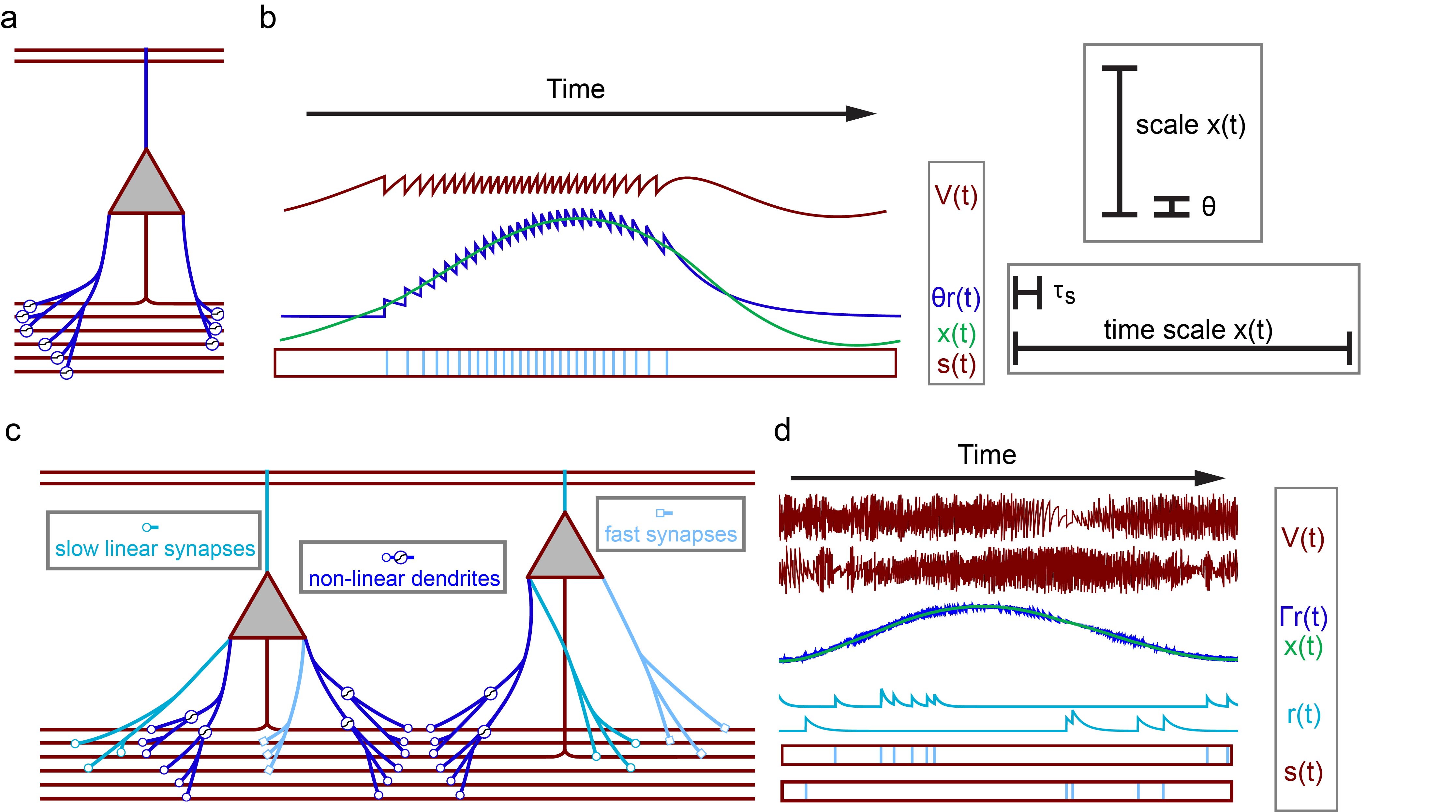}
\noindent \begin{centering}
\par\end{centering}
\protect\caption{
\textbf{Coding of continuous signals in neurons with saturating synapses (a,b)
and nonlinear dendrites (c,d).} (a,b): A neuron with saturating synapses
(a) that directly codes for a continuous signal (b). Panel (a) displays
the neuron with an axon (red) and dendrites (dark blue) that receive
inputs from the axons of other neurons (axons at the bottom) via saturating
synapses (symbolized by sigmoids at the synaptic contacts). The currents
entering the soma are weighted sums of input spike trains that are
synaptically filtered (generating scaled normalized synaptic currents
$\gamma r_{n}(t)$, synaptic time scale $\tau_{s}$) and thereafter
subject to a saturating synaptic nonlinearity. External inputs (axons
at the top) are received without saturation. The continuous signal
$x(t)$ (panel b left hand side, green) is the sum of the neuron's
membrane potential $V(t)$ (red) and its scaled normalized synaptic
current $\theta r(t)$ (dark blue). $r(t)$ is a low-pass filtered
version of the neuron's spike train $s(t)$ (light blue in red box).
If $x(t)>0$, the time scale of $x(t)$ should be large against the
synaptic time scale $\tau_{s}$ and $x(t)$ should predominantly be
large against the neuron's threshold, $\nicefrac{\theta}{2}$ (panel
b right hand side, assumptions {[}1,2{]} in the main text). $x(t)$
is then already well approximated by $\theta r(t)$, while $V(t)$
is oscillating between $\pm\nicefrac{\theta}{2}$. If $x(t)\leq0$,
we have $V(t)\leq0$, no spikes are generated and $r(t)$ quickly
decays to zero, such that we predominantly have $r(t)\approx0$ and
$x(t)$ is well approximated by $V(t)$ (cf.~Equation \eqref{eq:directapprox}).\protect \\
(c,d): Two neurons with nonlinear dendrites (c) from a larger network
that distributedly codes for a continuous signal (d). (c): Each neuron
has an axon (red) and different types of dendrites (cyan, light blue
and dark blue) that receive inputs from the axons of other neurons
(axons at the bottom) via fast or slow conventional synapses (highlighted
by circles and squares). Linear dendrites with slow synapses (cyan
with circle contacts) generate somatic currents that are weighted
linear sums of low-pass filtered presynaptic spike trains (weighted
sums of the $r_{n}(t)$). Linear dendrites with fast synapses (light
blue with square contacts) generate somatic currents with negligible
filtering (weighted sums of the spike trains $s_{n}(t)$). Spikes
arriving at a nonlinear dendrite (dark blue) are also filtered (circular
contact). The resulting $r_{n}(t)$ are weighted, summed up linearly
in the dendrite and subjected to a saturating dendritic nonlinearity
(symbolized by sigmoids at dendrites), before entering the soma. We
assume that the neurons have nonlinear dendrites that are located
in similar tissue areas, such that they connect to the same sets of
axons and receive similar inputs. (d): All neurons in the network
together encode $J$ continuous signals ${\bf x}(t)$ (one displayed
in green) by a weighted sum of their membrane potentials ${\bf V}(t)$
(two traces of different neurons displayed in red) and their normalized
PSCs ${\bf r}(t)$ (two traces displayed in cyan). The ${\bf \Gamma}{\bf r}(t)$
alone already approximate ${\bf x}(t)$ well. The neurons' output
spike trains ${\bf s}(t)$ (light blue in red box) generate slow and
fast inputs to other neurons. (Note that spikes can be generated due
to suprathreshold excitation by fast inputs. Since we plot ${\bf V}(t)$
after fast inputs and possible resets, the corresponding threshold
crossings do not appear.)
}
\end{figure}

The saturation of synapses, e.g.~due to receptor saturation or finite
reversal potentials, acts as a nonlinear transfer function \cite{blitz2004short,DA01},
which we model as a $\tanh$-nonlinearity (since $r_{m}(t)\geq0$
only the positive part of the $\tanh$ becomes effective). We note
that this may also be interpreted as a simple implementation of synaptic
depression: A spike generated by neuron $m$ at $t_{m}$ leads to
an increase of $r_{m}(t_{m})$ by $1$. As long as the synapse connecting
neuron $m$ to neuron $n$ is far from saturation (linear part of
the $\tanh$-function) this leads to the consumption of a fraction
$\gamma$ of the synaptic ``resources'' and the effect of the spike
on the neuron is approximately the effect of a current $A_{nm}\gamma e^{-\lambda_{s}(t-t_{m})}\Theta(t-t_{m})$.
When a larger number of such spikes arrive in short time such that
the consumed resources accumulate to $1$ and beyond, the synapse
saturates at its maximum strength $A_{nm}$ and the effect of individual
inputs is much smaller than before. The recovery from depression is
here comparably fast, it takes place on a timescale of $\lambda_{s}^{-1}$
(compare, e.g., \cite{AR04}).

The reset of the neuron is incorporated by the term $-\t s_{n}(t)$.
The voltage lost due to this reset is partially recovered by a slow
recovery current (afterdepolarization) $V_{\text{r}}\lambda_{s}r_{n}(t)$;
its temporally integrated size is given by the parameter $V_{\text{r}}$.
This is a feature of many neurons e.g.~in the neocortex, in the hippocampus
and in the cerebellum \cite{bean2007action}, and may be caused by
different types of somatic or dendritic currents, such as persistent
and resurgent sodium and calcium currents, or by excitatory autapses
\cite{luebke1996frequency,Bekkers:2009}. It provides a simple mechanism
to sustain (fast) spiking and generate bursts, e.g. in response to
pulses. $I_{\text{e,}n}(t)$ is an external input, its constant part
may be interpreted as sampling slow inputs specifying the resting
potential that the neuron asymptotically assumes for long times without
any recurrent network input. We assume that the resting potential
is halfway between the reset potential $V_{\text{res}}$ and the threshold
$V_{\text{res}}+\theta$. We set it to zero such that the neuron spikes
when the membrane potential reaches $\nicefrac{\theta}{2}$ and resets
to $\nicefrac{-\theta}{2}$. To test the robustness of the dynamics
we sometimes add a white noise input $\eta_{n}(t)$ satisfying $\left\langle \eta_{n}(t)\eta_{m}(t')\right\rangle =\sigma_{\eta}^{2}\delta_{nm}\delta(t-t')$
with the Kronecker delta $\delta_{nm}$.

For simplicity, we take the parameters $\lambda_{V},\,\t,V_{\text{r}}$
and $\gamma,\,\lambda_{s}$ identical for all neurons and synapses,
respectively. 
We take the membrane potential $V_{n}$ and the parameters
$V_{\text{r}}$ and $\t$ dimensionless, they can be fit to the voltage
scale of biological neurons by rescaling with an additive and a multiplicative
dimensionful constant. Time is measured in seconds.

We find that networks of the form Equation~\eqref{eq:Vdot} generate
dynamics suitable for universal computation similar to continuous
rate networks \cite{JH04,SA09}, if $0<\lambda_{x}\ll\lambda_{s}$,
where $\lambda_{x}=\lambda_{s}\left(1-\frac{V_{\text{r}}}{\t}\right)$,
$A_{nm}$ sufficiently large and $\gamma$ small. The conditions result
from requiring the network to approximate a nonlinear continuous dynamical
system (see next section).

An alternative interpretation of the introduced nonlinearity is that
the neurons have nonlinear dendrites, where each nonlinear compartment
is small such that it receives at most one (conventional, nonsaturating)
synapse. $A_{nm}$ is then the strength of the coupling from a dendritic
compartment to the soma. This interpretation suggests an extension
of the neuron model allowing for several dendrites per neuron, where
the inputs are linearly summed up and then subjected to a saturating
dendritic nonlinearity \cite{LH05,Memmesheimer2010,caze2013passive}.
Like the previous model, we find that such a model has to satisfy
additional constraints to be suitable for universal computation:

Neurons with nonlinear dendrites need additional slow and fast synaptic
contacts which arrive near the soma and are summed linearly there
(Fig.~1c). Such structuring has been found in biological neural networks
\cite{branco2011synaptic}. We gather the different components into
a dynamical equation for $V_{n}$ as 
\begin{align}
\dot{V}_{n}(t)= & -\lambda_{V}V_{n}(t)+\sum_{j=1}^{J}D_{nj}\tanh\left(\sum_{m=1}^{N}W_{njm}r_{m}(t)\right)\nonumber \\
 & +\sum_{m=1}^{N}\tilde{U}_{nm}r_{m}(t)-\sum_{m=1}^{N}U_{nm}s_{m}(t)\nonumber \\
 & +\sum_{j=1}^{J}\G_{jn}I_{\text{e},j}(t).\label{eq:VdotDistributed}
\end{align}
$D_{nj}$ is the coupling from the $j$th dendrite of neuron $n$
to its soma. The total number of dendrites and neurons is referred
to as $J$ and $N$ respectively. $W_{njm}$ is the coupling strength
from neuron $m$ to the $j$th nonlinear dendrite of neuron $n$.
The slow, significantly temporally filtered inputs from neuron $m$
to the soma of neuron $n$, $\tilde{U}_{nm}r_{m}(t)$, have connection
strengths $\tilde{U}_{nm}$. The fast ones, $U_{nm}s_{m}(t)$, have
negligible synaptic filtering (i.e.~negligible synaptic rise and
decay times) as well as negligible conduction delays. The resets and
recoveries are incorporated as diagonal elements of the matrices $U_{nm}$
and $\tilde{U}_{nm}$. To test the robustness of the dynamics, also
here we sometimes add a white noise input $\eta_{n}(t)$. To increase
the richness of the recurrent dynamics and the computational power
of the network (cf.~\cite{huang2006extreme} for disconnected units
without output feedback) we added inhomogeneity, e.g.~through the
external input current in some tasks. In the control/mental exploration task, we added
a constant bias term $b_{j}$ as argument of the $\tanh$ to introduce
inhomogeneity.

We find that the network couplings $\mathbf{D},\mathbf{W},\mathbf{U}$
and $\tilde{\mathbf{U}}$ \eqref{eq:VdotDistributed} (we use bold
letters for vectors and matrices) should satisfy certain interrelations.
As motivated in the subsequent section and derived in the supporting
material, their components may be expressed in terms of the components
of a $J\times N$ matrix $\mathbf{\Gamma}$, and a $J\times J$ matrix
$\mathbf{A}$ as $D_{nj}=\sum_{i=1}^{J}\Gamma_{in}A_{ij}$, $W_{njm}=\Gamma_{jm}$,
$\tilde{U}_{nm}=a\sum_{j=1}^{J}\G_{jn}\G_{jm}+\mu\lambda_{s}\delta_{nm}$,
$U_{nm}=\sum_{j=1}^{J}\G_{jn}\G_{jm}+\mu\delta_{nm}$, where $a=\lambda_{s}-\lambda_{x}$
and $\mu\geq0$ is small (see also Table 1 for an overview). The thresholds are chosen identical,
$\theta^n=\theta$, see Methods.
\begin{table}
\textbf{}%
\begin{tabular}{|l||c|c|}
\hline 
 & \textbf{explanation} & \textbf{optimal value}\tabularnewline
\hline 
\textbf{$D_{nj}$ } & coupling from the $j$th dendrite & \textbf{$D_{nj}=\sum_{i=1}^{J}\Gamma_{in}A_{ij}$}\tabularnewline
 & of neuron $n$to its soma &
\tabularnewline
\hline 
\textbf{$W_{njm}$ } & coupling strength from neuron $m$ to & \textbf{$W_{njm}=\Gamma_{jm}$ }\tabularnewline
 & the $j$th nonlinear dendrite of neuron $n$ & 
\tabularnewline
\hline 
\textbf{$\tilde{U}_{nm}$ } & slow coupling from neuron $m$ to neuron $n$; & \textbf{$\tilde{U}_{nm}=a\sum_{j=1}^{J}\G_{jn}\G_{jm}+\mu\lambda_{s}\delta_{nm}$}\tabularnewline
 & diagonal elements incorporate & \textbf{$a=\lambda_{s}-\lambda_{x}$}\tabularnewline
 & a recovery current &
\tabularnewline
\hline 
\textbf{$U_{nm}$ } & fast coupling from neuron $m$ to neuron $n$; & \textbf{$U_{nm}=\sum_{j=1}^{J}\G_{jn}\G_{jm}+\mu\delta_{nm}$ }\tabularnewline
 & diagonal elements incorporate the reset &
\tabularnewline
\hline 
\textbf{$\theta^{n}$ } & threshold of neuron $n$ & \textbf{$\theta^{n}=\frac{U_{nn}}{2}$ }\tabularnewline
\hline 
\end{tabular}\textbf{\protect\caption{Parameters of a network of neurons with nonlinear dendrites (cf.~Equation~\eqref{eq:VdotDistributed})
and their optimal values.\textbf{\label{tab:NonlinearDendrites}}}
}
\end{table}

Again, the conditions result from requiring the network to approximate
a nonlinear continuous dynamical system. This system, Equation~\eqref{eq:xdotDistributed},
is characterized by the $J\times J$ coupling matrix $\mathbf{A}$
and a $J$-dimensional input ${\bf c}(t)$ whose components are identical
to the $J$ independent components of the external input current $\mathbf{I}_{\text{e}}$
in equation~\eqref{eq:VdotDistributed}; the matrix $\mathbf{\Gamma}$
is a decoding matrix that fixes the relation between spiking and continuous
dynamics (see next section). We note that the matrices $\mathbf{\Gamma}$
and $\mathbf{A}$ are largely unconstrained, such
that the coupling strengths maintain a large degree of arbitrariness.
Ideally, $W_{njm}$ is independent of $n$, therefore neurons have
dendrites that are similar in their input characteristics to dendrites
in some other neurons (note that $\mathbf{D}$ may have zero entries,
so dendrites can be absent). We interpret these as dendrites that
are located in a similar tissue area and therefore connect to the
same axons and receive similar inputs (cf.~Fig.~1c for an illustration).
The interrelations between the coupling matrices might be realized
by spike-timing dependent synaptic or structural plasticity. Indeed,
for a simpler model and task, appropriate biologically plausible learning
rules have been recently highlighted \cite{bourdoukan2012learning,Bourdoukan2015}.
We tested robustness of our schemes against structural perturbations
(see Figs.~C, D in S1 Supporting Information), in particular for deviations from the $n$-independence
of $W_{njm}$ (Fig.~C in S1 Supporting Information).

The networks Equation~\eqref{eq:Vdot} with saturating synapses have
a largely unconstrained topology, in particular they can satisfy the
rule that neurons usually act only excitatorily or inhibitorily. For
the networks Equation~\eqref{eq:VdotDistributed} with nonlinear
dendrites, it is less obvious how to reconcile the rule with the constraints
on the network connectivity. Solutions for this have been suggested
in simpler systems and are subject to current research \cite{boerlin2013predictive}.

The key property of the introduced neural architecture is that the
spike trains generated by the neurons encode with high signal-to-noise
ratio a continuous signal that can be understood in terms of ordinary
differential equations. In the following section we show how this
signal is decoded from the spike trains. Thereafter, we may conclude
that the spiking dynamics are sufficiently ``tamed'' such that standard
learning rules can be applied to learn complicated computations.

\subsubsection*{Direct encoding of continuous dynamics}

The dynamics of a neural network with $N$ integrate-and-fire neurons
consist of two components, the sub-threshold dynamics $\mathbf{V}(t)=(V_{1}(t),...,V_{N}(t))^{T}$
of the membrane potentials and the spike trains $\mathbf{s}(t)=(s_{1}(t),...,s_{N}(t))^{T}$
(Equation~\ref{eq:st}), which are temporal sequences of $\delta$-distributions.
In the model with saturating synapses, all synaptic interactions are
assumed to be significantly temporally filtered, such that the $V_{n}(t)$
are continuous except at reset times after spiking (Equation \eqref{eq:Vdot}).
We posit that the $\mathbf{V}(t)$ and the $\mathbf{s}(t)$ should
together form some $N$-dimensional continuous dynamics $\mathbf{x}(t)=(x_{1}(t),...,x_{N}(t))^{T}$.
The simplest approach is to setup $\mathbf{x}(t)$ as a linear combination
of the two components $\mathbf{V}(t)$ and $\mathbf{s}(t)$. To avoid
infinities in $x_{n}(t)$, we need to eliminate the occurring $\delta$-distributions
by employing a smoothed version of $s_{n}(t)$. This should have a
finite discontinuity at spike times such that the discontinuity in
$V_{n}(t)$ can be balanced. A straightforward choice is to use $\t r_{n}(t)$
(Equation~\eqref{eq:rt}) and to set
\begin{equation}
V_{n}(t)+\t r_{n}(t)=x_{n}(t).\label{eq:Vrx}
\end{equation}
(cf.~Fig.~1b). When the abovementioned
conditions on $\lambda_{x}$, $\lambda_{s}$,
$\mathbf{A}$ and $\gamma$ are satisfied (cf.~end of the section
introducing networks with saturating synapses), the continuous signal
$\mathbf{x}(t)$ follows a system of first order nonlinear ordinary
differential equations similar to those
describing standard non-spiking continuous rate networks used for
computations (cf.~\cite{JH04,SA09,lukosevicius2012reservoir} and
Equation \eqref{eq:xdotDistributed} below),
\begin{align}
\dot{x}_{n}(t)= & -\lambda_{V}\left[x_{n}(t)\right]_{-}-\lambda_{x}\left[x_{n}(t)\right]_{+}+\sum_{m=1}^{N}A_{nm}\tanh\left(\frac{\gamma}{\theta}\left[x_{m}(t)\right]_{+}\right)+I_{\text{e},n}(t),\label{eq:xdot}
\end{align}
with the rectifications $\left[x_{n}(t)\right]_{+}=\max\left(x_{n}(t),0\right),$
$\left[x_{n}(t)\right]_{-}=\min\left(x_{n}(t),0\right)$. We call
spiking networks where this is the case \emph{continuous signal coding
spiking neural networks (CSNs)}.

Except for the rectifications, Equation \eqref{eq:xdot} has a standard
form for non-spiking continuous rate networks, used for computations
\cite{JH04,SA09,lukosevicius2012reservoir}. A
salient choice for $\lambda_{x}$ is $\lambda_{x}=\lambda_{V}$, i.e.~$V_{\text{r}}=\left(1-\frac{\lambda_{V}}{\lambda_{s}}\right)\theta$,
such that the rectifications outside the $\tanh$-nonlinearity vanish.
Equation \eqref{eq:xdot} generates dynamics that are different from
the standard ones in the respect that the trajectories of individual
neurons are, e.g.~for random Gaussian matrices $\mathbf{A}$, not
centered at zero. However, they can satisfy the conditions for universal
computation (enslaveability/echo state property and high dimensional
nonlinear dynamics) and generate longer-term fading memory for appropriate
scaling of $\mathbf{A}$. Also the corresponding spiking networks
are then suitable for fading memory-dependent computations. Like
for the standard networks \cite{jaeger2001echo,yildiz2012re}, we
can derive sufficient conditions to guarantee that the dynamics Equation
\eqref{eq:xdot} are enslaveable by external signals (echo state property).
$\left\Vert {\bf A}\right\Vert <\min(\lambda_{V},\lambda_{x})$, where
$\left\Vert {\bf A}\right\Vert $ is the largest singular value of
the matrix ${\bf A}$, provides such a condition (see Supplementary
material for the proof). The condition is rather strict, our applications
indicate that the CSNs are also suited as computational reservoirs
when it is violated. This is similar to the situation in standard
rate network models \cite{jaeger2001echo}. We note that if the system
is enslaved by an external signal, the time scale of $x_{n}(t)$ is
largely determined by this signal and not anymore by the intrinsic
scales of the dynamical system.

We will now show that spiking neural networks Equation \eqref{eq:Vdot}
can encode continuous dynamics Equation \eqref{eq:xdot}. For this
we derive the dynamical equation of the membrane potential \eqref{eq:Vdot}
from the dynamics of ${\bf x}(t)$ using the coding rule Equation~\eqref{eq:Vrx},
the dynamical equation \eqref{eq:rt} for $r_{n}(t)$ and the rule
that a spike is generated whenever $V_{n}(t)$ reaches threshold $\nicefrac{\theta}{2}$:
We first differentiate Equation \eqref{eq:Vrx} to eliminate $\dot{x}_{n}(t)$
from Equation \eqref{eq:xdot} and employ Equation \eqref{eq:rt}
to eliminate $\dot{r}_{n}(t)$. The resulting expression for $\dot{V}_{n}(t)$
reads
\begin{equation}
\dot{V}_{n}(t)=-\lambda_{V}\left[x_{n}(t)\right]_{-}-\lambda_{x}\left[x_{n}(t)\right]_{+}+\sum_{m=1}^{N}A_{nm}\tanh\left(\frac{\gamma}{\theta}\left[x_{m}(t)\right]_{+}\right)-\theta s_{n}(t)+\lambda_{s}\theta r_{n}(t)+I_{\text{e,}n}(t).\label{eq:xdot-2}
\end{equation}
It already incorporates the resets of size $\theta$ (cf.~the term
$-\theta s_{n}(t)$), they arise since $x_{n}(t)=V_{n}(t)+\theta r_{n}(t)$
is continuous and $r_{n}(t)$ increases by one at spike times (thus
V must decrease by $\theta$). We now eliminate the occurrences of
$\left[x_{n}(t)\right]_{+}$ and $\left[x_{n}(t)\right]_{-}$.

For this, we make two assumptions (cf\@.~Fig.~1b)
on the $x_{n}(t)$ if they are positive:
\begin{enumerate}
\item[{{[}1{]} }]  The dynamics of $x_{n}(t)$ are slow
against the synaptic timescale $\tau_{s}$,
\item[{{[}2{]} }]  the $x_{n}(t)$ assume predominantly
values $x_{n}(t)\gg\nicefrac{\theta}{2}$.
\end{enumerate}
First we consider the case $x_{n}(t)>0$.
Since $V_{n}(t)$ is reset when it reaches its threshold value $\nicefrac{\theta}{2}$,
$V_{n}(t)$ is always smaller than $\nicefrac{\theta}{2}$. Thus,
given $V_{n}(t)>0$ assumption {[}2{]}
implies that we can approximate $x_{n}(t)\approx\theta r_{n}(t)$,
as the contribution of $V_{n}(t)$ is negligible because $V_{n}(t)\leq\nicefrac{\theta}{2}$.
This still holds if $V_{n}(t)$ is negative and its absolute value
is not large against $\nicefrac{\theta}{2}$. Furthermore, assumption
{[}1{]} implies that smaller negative $V_{n}(t)$ cannot co-occur
with positive $x_{n}(t)$: $r_{n}(t)$ is positive and in the absence
of spikes it decays to zero on the synaptic time scale $\tau_{s}$
(Equation~\eqref{eq:rt}).
When $V_{n}(t)<0$, neuron $n$ is not spiking anymore. Thus when
$V_{n}(t)$ is shrinking towards small negative values and $r_{n}(t)$
is decaying on a timescale of $\tau_{s}$, $x_{n}(t)$ is also decaying
on a time-scale $\tau_{s}$. This contradicts assumption {[}1{]}.
Thus when $x_{n}(t)>0$, the absolute magnitude
of $V_{n}(t)$ is on the order of $\nicefrac{\theta}{2}$. With assumption
{[}2{]} we can thus set $x_{n}(t)\approx\theta r_{n}(t)$, whenever
$x_{n}(t)>0$, neglecting contributions of size $\nicefrac{\theta}{2}$.

Now we consider $x_{n}(t)\leq0$. This implies $V_{n}(t)\leq0$ (since
always $r_{n}(t)\geq0$) as well as a quick decay of $r_{n}(t)$ to
zero. When $x_{n}(t)$ assumes values significantly below zero, assumption
{[}1{]} implies that we have $x_{n}(t)\approx V_{n}(t)$ and $r_{n}(t)\approx0$,
otherwise $x_{n}(t)$ must have changed from larger positive (assumption
{[}2{]}) to larger negative values on a timescale of $\tau_{s}$.

The approximate expressions may be gathered in the replacements $\left[x_{n}(t)\right]_{+}=\theta r_{n}(t)$
and $\left[x_{n}(t)\right]_{-}=\left[V_{n}(t)\right]_{-}$. Using
these in Equation \eqref{eq:xdot-2} yields together with $\lambda_{x}=\lambda_{s}\left(1-\frac{V_{\text{r}}}{\t}\right)$
\begin{equation}
\dot{V}_{n}(t)=-\lambda_{V}\left[V_{n}(t)\right]_{-}+\sum_{m=1}^{N}A_{nm}\tanh\left(\gamma r_{m}(t)\right)+V_{\text{r}}\lambda_{s}r_{n}(t)-\theta s_{n}(t)+I_{\text{e,}n}(t).		\label{eq:xdot-3}
\end{equation}
Note that our replacements allowed to eliminate the biologically implausible
${\bf V}$-dependencies in the interaction term.

To simplify the remaining $V_{n}(t)$-dependence, we additionally
assume that
\begin{enumerate}
\item[{{[}2'{]} }] $x_{n}(t)$ assumes predominantly values
$x_{n}(t)\gg\lambda_{V}\theta/(2\lambda_{x})$,
\end{enumerate}
if $x_{n}(t)$ is positive. This can be
stricter than {[}2{]} depending on the values of $\lambda_{x}$ and
$\lambda_{V}$. For positive $x_{n}(t)$, where $\lambda_{V}\left[x_{n}(t)\right]_{-}$
in Equation \eqref{eq:xdot-2} is zero, $\lambda_{V}V_{n}(t)$
has an absolute magnitude on the order of $\lambda_{V}\theta/2$ (see
the arguments above). {[}2'{]} implies that this
is negligible against $-\lambda_{x}\left[x_{n}(t)\right]_{+}$. For
negative $x_{n}(t)$, we still have $x_{n}(t)\approx V_{n}(t)$. This
means that we may replace $-\lambda_{V}\left[x_{n}(t)\right]_{-}$
by $\lambda_{V}V_{n}(t)$ in Equation \eqref{eq:xdot-2}. Taken together,
under the assumptions {[}1,2,2'{]} we may use the replacements 
\begin{align}
\left[x_{n}(t)\right]_{+} & \approx\theta r_{n}(t)\nonumber \\
\left[x_{n}(t)\right]_{-} & \approx V_{n}(t)\label{eq:directapprox}
\end{align}
in Equation \eqref{eq:xdot-2}, which directly yield Equation \eqref{eq:Vdot}.
Note that this also implies $r_{n}(t)\gg\nicefrac{\theta}{2}$
if the neuron is spiking, so during active periods inter-spike-intervals
need to be considerably smaller than the synaptic time scale.

Equation \eqref{eq:xdot} implies that the assumptions are justified
for suitable parameters: For fixed parameters $\tau_{s}=\lambda_{s}^{-1}$
and $\theta$ of the ${\bf r}$-dynamics, we can choose sufficiently
small $\lambda_{x}$, large $A_{nm}$ and
small $\gamma$ to ensure assumptions {[}1,2,2'{]} (cf.~the conditions
highlighted in the section ``Network architecture''). On the other
hand, for given dynamics Equation \eqref{eq:xdot}, we can always
find a spiking system which generates the dynamics via Equations \eqref{eq:Vdot},
\eqref{eq:rt} and \eqref{eq:Vrx}, and satisfies the assumptions:
We only need to choose $\tau_{s}$ sufficiently small such that {[}1{]}
is satisfied and the spike threshold sufficiently small such that
{[}2,2'{]} are satisfied. For the latter, $\gamma$ needs to be scaled
like $\theta$ to maintain the dynamics of $x_{n}$ and $V_{r}$ needs
to be computed from the expression for $\lambda_{x}$. Interestingly,
we find that also outside the range where the assumptions are satisfied,
our approaches can still generate good results.

The recovery current in our model has the same time
constant as the slow synaptic current. Indeed, experiments indicate
that they possess the same characteristic timescales: Timescales for
NMDA \cite{wang2008specialized} and slow GABA$_{\text{A}}$ \cite{petrides2007gabaa,sceniak2008slow}
receptor mediated currents are several tens of milliseconds. Afterdepolarizations
have timescales of several tens of milliseconds as well \cite{storm1987action,raman1997resurgent,bean2007action,chen2008spike,brown2009activity}.
Another prominent class of slow inhibitory currents is mediated by
GABA$_{\text{B}}$ receptors and has time scales of one hundred to
a few hundreds of milliseconds \cite{luescher1997gprotein}. We remark
that in our model the time constants of the afterdepolarization and
the synaptic input currents may also be different without changing
the dynamics: Assume that the synaptic time constant is different
from that of the recovery current, but still satisfies the conditions
that it is large against the inter-spike-intervals when the neuron
is spiking and small against the timescale of $\left[x_{n}(t)\right]_{+}$.
The synaptic current generated by the spike train of neuron $n$ will
then be approximately continuous and the filtering does not seriously
affect its overall shape beyond smoothing out the spikes. As a consequence,
the synaptic and the recovery currents are approximately proportional
up to a constant factor that results from the different integrated
contribution of individual spikes to them. Rescaling $\gamma$ by
this factor thus yields dynamics equivalent to the one with identical
time constants.

\subsubsection*{Distributed encoding of continuous dynamics}

In the above-described simple CSNs (CSNs with saturating synapses),
each spiking neuron gives rise to one nonlinear continuous variable.
The resulting condition that the inter-spike-intervals are small against the synaptic time constants if the neuron
is spiking may in biological neural networks be satisfied for bursting
or fast spiking neurons with slow synaptic currents. It will be invalid
for different neurons and synaptic currents. The condition becomes
unnecessary when the spiking neurons encode continuous variables collectively,
i.e.~if we partially replace the temporal averaging in $r_{n}(t)$
by an ensemble averaging. This can be realized by an extension of
the above model, where only a lower, say $J-$, dimensional combination
$\mathbf{x}(t)$ of the $N-$dimensional
vectors $\mathbf{V}(t)$ and $\mathbf{r}(t)$ is continuous,
\begin{equation}
\mathbf{x}(t)=\mathbf{L}\mathbf{V}(t)+\mathbf{\tilde{\Gamma}}\mathbf{r}(t),\label{eq:xLvGx}
\end{equation}
where $\mathbf{L}$ and $\mathbf{\tilde{\mathbf{\Gamma}}}$ are $J\times N$
matrices (note that Equation \eqref{eq:Vrx} is a special case with
$N=J$ and diagonal matrices $\mathbf{L}$ and $\mathbf{\tilde{\mathbf{\Gamma}}}$).
We find that spiking networks with nonlinear dendrites Equation \eqref{eq:VdotDistributed}
can encode such a lower dimensional variable $\mathbf{x}(t)$.
The $\mathbf{x}(t)$ satisfy $J$-dimensional
standard equations describing non-spiking continuous rate networks
used for reservoir computing \cite{JH04,SA09,lukosevicius2012reservoir},
\begin{equation}
\dot{\mathbf{x}}(t)=-\lambda_{x}\mathbf{x}(t)+\mathbf{A}\tanh\left(\mathbf{x}(t)\right)+\mathbf{c}(t).\label{eq:xdotDistributed}
\end{equation}
We denote the resulting spiking networks as CSNs with nonlinear dendrites.

The derivation (see Supplementary material for details) generalizes
the ideas introduced in refs.~\cite{boerlin2011spike,bourdoukan2012learning,boerlin2013predictive}
to the approximation of nonlinear dynamical systems: We assume an
approximate decoding equation (cf.~also Equation \eqref{eq:directapprox}),
\begin{equation}
\mathbf{x}(t)\approx\mathbf{\Gamma}\mathbf{r}(t),\label{eq:distributedapprox}
\end{equation}
where $\mathbf{\mathbf{\boldsymbol{\Gamma}}}$ is a $J\times N$ decoding
matrix and employ an optimization scheme that minimizes the decoding
error resulting from Equation \eqref{eq:distributedapprox} at each
time point. This yields the condition that a spike should be generated
when a linear combination of ${\bf x}(t)$ and ${\bf r}(t)$ exceeds
some constant value. We interpret this linear combination as membrane
potential ${\bf V}(t)$. Solving for ${\bf x}(t)$ gives $\mathbf{L}$
and $\tilde{\mathbf{\Gamma}}$ in terms of $\mathbf{\Gamma}$ in Equation
\eqref{eq:xLvGx}. Taking the temporal derivative yields $\dot{{\bf V}}(t)$,
first in terms of $\dot{{\bf x}}(t)$ and $\dot{{\bf r}}(t)$ and
after replacing them via Equations \eqref{eq:rt},\eqref{eq:xdotDistributed},
in terms of ${\bf x}(t)$, ${\bf r}(t)$ and ${\bf s}(t)$. We then
eliminate ${\bf x}(t)$ using \eqref{eq:distributedapprox} and add
a membrane potential leak term for biological realism and increased
stability of numerical simulations. This yields Equation \eqref{eq:VdotDistributed}
together with the optimal values of the parameters given in Table
1. We note that the difference to the derivation in ref.~\cite{boerlin2013predictive}
is the use of a nonlinear equation when replacing $\dot{{\bf x}}(t)$.
We further note that the spiking approximation of the continuous dynamics
becomes exact, if in the last step ${\bf x}(t)$ is eliminated using
Equation \eqref{eq:xLvGx} and the leak term is omitted as it does
not arise from the formalism in contrast to the case of CSNs with
saturating synapses. Like in CSNs with saturating synapses, using
the approximated decoding Equation \eqref{eq:distributedapprox} eliminates
the biologically implausible ${\bf V}$-dependencies in the interaction
terms. For an illustration of this coding see Fig.~1d.

\subsection*{Learning universal computations}

Recurrent continuous rate networks are a powerful means for learning
of various kinds of computations, like steering of movements and processing
of sequences \cite{JH04,SA09}. For this, an input and/or an output
feedback signal needs to be able to ``enslave'' the network's high-dimensional
dynamics \cite{jaeger2001echo,yildiz2012re}. This means that at any
point in time the network's state is a deterministic function of the
recent history of input and feedback signals. The function needs to
be high dimensional, nonlinear, and possess fading memory. A standard
model generating suitable dynamics are continuous rate networks of
the form Equation~\eqref{eq:xdotDistributed}. Due to the typically
assumed random recurrent connectivity, each neuron acts as a randomly
chosen, nonlinear function with fading memory. Linearly combining
them like basis functions by a linear readout can approximate arbitrary,
nonlinear functions with fading memory (time-scales are limited by
the memory of the network), and in this sense \emph{universal computations}
on the input and the feedback. The feedback can prolong the fading
memory and allow to generate self-contained dynamical systems and
output sequences \cite{JH04,maass2007computational,SA09,SA12}. The
feedback can be incorporated into the network by directly training
the recurrent synaptic weights \cite{SA09,SA12}.

Our understanding of the complex spiking dynamics of CSNs in terms
of nonlinear first order differential equations enables us to apply
the above theory to spiking neural networks: In the first step, we
were able to conclude that our CSNs can generate enslaveable and thus
computationally useful dynamics as they can be decoded to continuous
dynamics that possess this property. In the second step, we have to
ask which and how output signals should be learned to match a desired
signal: In a biological setting, the appropriate signals are the sums
of synaptic or dendritic input currents that spike trains generate,
since these affect the somata of postsynaptic neurons as well as effectors
such as muscles \cite{eliasmith2003neural}. To perform, e.g., a desired
continuous movement, they have to prescribe the appropriate muscle
contraction strengths. For both CSNs with saturating synapses and
with nonlinear dendrites, we choose the outputs to have the same form
as the recurrent inputs that a soma of a neuron within the CSN receives.
Accordingly, in our CSNs with saturating synapses, we interpret sums
of the postsynaptic currents 
\begin{equation}
z_{k}(t)=\sum_{m=1}^{N}w_{km}^{o}\tanh\left(\gamma r_{m}(t)\right)=:\sum_{m=1}^{N}w_{km}^{o}\widetilde{r}_{m}(t)\label{eq:zk saturating synapses}
\end{equation}
as output signals, where the index $k$ distinguishes $K_{\text{out}}$
different outputs, and $w_{km}^{o}$ are the learnable synaptic output
weights. For networks with nonlinear dendrites the outputs are a linear
combination of inputs preprocessed by nonlinear dendrites 
\begin{equation}
z_{k}(t)=\sum_{j=1}^{J}w_{kj}^{o}\tanh\left(\sum_{m=1}^{N}\Gamma_{jm}r_{m}(t)\right)=:\sum_{j=1}^{J}w_{kj}^{o}\widetilde{r}_{j}(t),\label{eq:zk nonlinear dendrites}
\end{equation}
where the strengths $w_{kj}^{o}$ of the dendro-somatic coupling are
learned \cite{LMM08}. The networks can now learn the output weights
such that $z_{k}(t)$ imitates a target signal $F_{k}(t)$, using
standard learning rules for linear readouts (see Fig.~2a for an illustration).
We employ the recursive least squares method \cite{HS02}.

To increase the computational and learning abilities, the output signals
should be fed back to the network as an (additional) input (Fig.~2b)
\begin{equation}
I_{\text{e},\beta}^{f}(t)=\sum_{k=1}^{K_{\text{out}}}w_{\beta k}^{f}z_{k}(t)=\sum_{k=1}^{K_{\text{out}}}w_{\beta k}^{f}\sum_{\rho}w_{k\rho}^{o}\widetilde{r}_{\rho}(t),\label{eq:feedback-1-1}
\end{equation}
where each neuron receives a linear combination of the output signals
$z_{k}(t)$ with static feedback connection strengths $w_{\beta k}^{f}$.
Here and in the following Greek letter indices such as $\beta,\rho$
range over all saturating synapses ($\beta,\rho=1,...,N$; $\tilde{r}_{\beta}(t)=\tanh\left(\gamma r_{\beta}(t)\right)$)
in CSNs with saturating synapses, or over all nonlinear dendrites
($\beta,\rho=1,...,J$; $\tilde{r}_{\beta}(t)=\tanh\left(\sum_{m=1}^{N}\Gamma_{\beta m}r_{m}(t)\right)$)
in CSNs with nonlinear dendrites.

It often seems biologically more plausible not to assume a strong
feedback loop that enslaves the recurrent network, but rather to train
recurrent weights. Our CSNs allow for this (Fig.~2c): We can transform
the learning of output weights in networks with feedback into mathematically
equivalent learning of recurrent connection strengths, between synapses
(CSNs with saturating synapses) or dendrites (CSNs with nonlinear
dendrites) and the soma \cite{LMM08} (we learn $A_{nm}$, see Methods
for details of the implementation). We note that approximating different
dynamical systems, e.g.~ones equivalent to Equation \eqref{eq:xdotDistributed}
but with the coupling matrix inside the nonlinearity \cite{hirsch1989convergent},
may also in CSNs with nonlinear dendrites allow to learn synaptic
weights in similar manner. We call CSNs with learning of outputs in
presence of feedback, or with learning of recurrent connections \emph{plastic
continuous signal coding spiking neural networks (PCSNs).}

To learn feedback and recurrent connections, we use the FORCE imitation
learning rule, which has recently been suggested for networks of continuous
rate neurons \cite{SA09,SA12}: We use fast online learning based
on the recursive least squares rule of the output weights in order
to ensure that the output of the network is similar to the desired
output at all times. Since during training the output is ensured to
be close to the desired one, it can be used as feedback to the network
at all times. The remaining deviations from the desired output are
expected to be particularly suited as training noise as they reflect
the system's inherent noise. As mentioned before, the feedback loop
may be incorporated in the recurrent network connectivity. During
training, the reservoir connections are then learned in a similar
manner as the readout.

In the following, we show that our approach allows spiking neural
networks to perform a broad variety of tasks. In particular, we show
learning of desired self-sustained dynamics at a degree of difficulty
that has, to our knowledge, previously only been accessible with continuous
rate networks.

\begin{figure}
\noindent \begin{centering}
\includegraphics[width=1\textwidth]{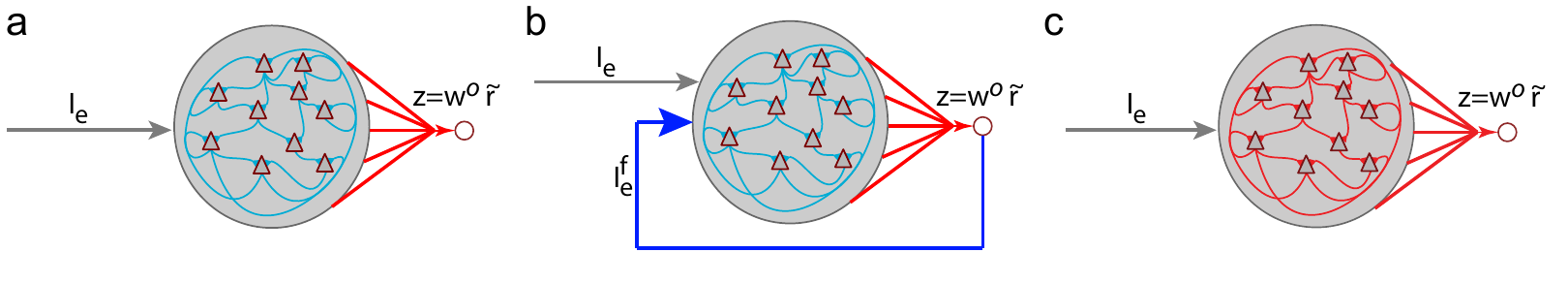}
\par\end{centering}

\protect\caption{
\textbf{Setups used to learn versatile nonlinear computations with spiking
neural networks.} (a) A static \emph{continuous
signal coding spiking neural network} (\emph{CSN},
gray shaded) serves as a spiking computational
reservoir with high signal-to-noise ratio. The results of computations
on current and past external inputs ${\bf I}_{\text{e}}$ can be extracted
by simple neuron-like readouts. These linearly combine somatic inputs
generated by saturating synapses or nonlinear dendrites, $\tilde{{\bf r}}$
(red), to output signals ${\bf z}$ (Equations (\ref{eq:zk saturating synapses},
\ref{eq:zk nonlinear dendrites})). The output weights ${\bf w}^{\text{o}}$
are learned such that ${\bf z}$ approximates the desired continuous
target signals. (b) \emph{Plastic continuous signal coding spiking
neural networks (PCSNs)} possess a loop that feeds the outputs ${\bf z}$
back via static connections as an additional input ${\bf I}_{\text{e}}^{f}$
(blue, Equation \ref{eq:feedback-1-1}). Such networks have increased
computational capabilities allowing them to, e.g., generate desired
self-sustained activity. (c) The feedback loop can be incorporated
into the recurrent network via plastic recurrent connections (red
in gray shaded area).
}
\end{figure}

\subsection*{Applications}

\subsubsection*{Self-sustained pattern generation}

Animals including humans can learn a great variety of movements, from
periodic patterns like gait or swimming, to much more complex ones
like producing speech, generating chaotic locomotion \cite{G06,li2014chaos}
or playing the piano. Moreover when an animal learns to use an object
(Fig.~3a), it has to learn the dynamical properties of the object
as well as how its body behaves when interacting with it. Especially
for complex, non-periodic dynamics, a dynamical system has to be learned
with high precision.

How are spiking neural networks able to learn dynamical systems, store
them and replay their activity? We find that PCSNs may solve the problem.
They are able to learn periodic patterns of different degree of complexity
as well as chaotic dynamical systems by imitation learning. Fig.~3
illustrates this for PCSNs with nonlinear synapses (Fig.~3d,e) and
with nonlinear dendrites (Fig.~3b,e,f).

The figure displays the recall of three periodic movements after learning:
a sine wave, a more complicated non-differentiable saw tooth pattern
and a ``camel's hump'' superposition of sine and cosine. Also for
long simulation times, we find no deviation from the displayed dynamics
except for an inevitable phase shift (Fig.~Ga in S1 Supporting Information). It results from
accumulation of small differences between the learned and desired
periods. Apart from this, the error between the recalled and the desired
signals is approximately constant over time (Fig.~Gb in S1 Supporting Information). This indicates
that the network has learned a stable periodic orbit to generate the
desired dynamics, the orbit is sufficiently stable to withstand the
intrinsic noise of the system. Fig.~3 furthermore illustrates learning
of a chaotic dynamical system. Here, the network learns to generate
the time varying dynamics of all three components of the Lorenz system
and produces the characteristic attractor pattern after learning (Fig.~3f).
Due to the encoding of the dynamics in spike trains, the signal maintains
a small deterministic error which emerges from the encoding of a continuous
signal by discrete spikes (Fig.~3g). The individual training and
recall trajectories quickly depart from each other after the end of
learning since they are chaotic. However, also for long simulation
times, we observe qualitatively the same dynamics, indicating that
the correct dynamical system was learned (Fig.~Gc in S1 Supporting Information). Occasionally,
errors occur, cf.~the larger loop in Fig.~3f. This is to be expected
due to the relatively short training period, during which only a part
of the phase space covered by the attractor is visited. Importantly,
we observe that after errors the dynamics return to the desired ones
indicating that the general stability property of the attractor is
captured by the learned system. To further test these observations,
we considered a not explicitly trained long-term feature of the Lorenz-dynamics,
namely the tent-map which relates the height $z_{n-1}$ of the $(n-1)$th
local maximum in the $z-$coordinate, to the height $z_{n}$ of the
subsequent local maximum. The spiking network indeed generates the
map (Fig.~3h), with two outlier points corresponding to each error.

In networks with saturating synapses, the spike trains are characterized
by possibly intermittent periods of rather high-frequency spiking.
In networks with nonlinear dendrites, the spike trains can have low
frequencies and they are highly irregular (Figs.~3c, F in S1 Supporting Information). In agreement
with experimental observations (e.g.\ \cite{quiroga2005invariant}),
the neurons can have preferred parts of the encoded signal in which
they spike with increased rates.

The dynamics of the PCSNs and the generation of the desired signal
are robust against dynamic and structural perturbations. They sustain
noise inputs which would accumulate to several ten percent of the
level of the threshold within the membrane time constant, for a neuron
without further input (Fig.~B in S1 Supporting Information). For larger deviations
of $W_{njm}$ from their optimal values, PCSNs with nonlinear dendrites
can keep their learning capabilities, if $\mu$ is tuned to a specific
range. Outside this range, the capabilities break down at small deviations
(Fig.~C in S1 Supporting Information). However, a slightly modified version of the models, where
the reset is always to $-\theta$ (even if there was fast excitation
that drove the neuron to spike by a suprathreshold input), has a
high degree of robustness against such structural perturbations. We
also checked that the fast connections are important, albeit substantial
weakening can be tolerated (Fig.~D in S1 Supporting Information).

The deterministic spike code of our PCSNs encodes the output signal
much more precisely than neurons generating a simple Poisson code,
which facilitates learning. We have quantified this using a comparison
between PCSNs with saturating synapses and networks of Poisson neurons
of equal size, both learning the saw tooth pattern in the same manner.
Since both codes become more precise with increasing spike rate of
individual neurons, we compared the testing error between networks
with equal spike rates. Due to their higher signal-to-noise ratio,
firing rates required by the PCSNs to achieve the same pattern generation
quality are more than one order of magnitude lower (Fig.~A in S1 Supporting Information).

\begin{figure}
\noindent \begin{centering}
\includegraphics[width=1\textwidth]{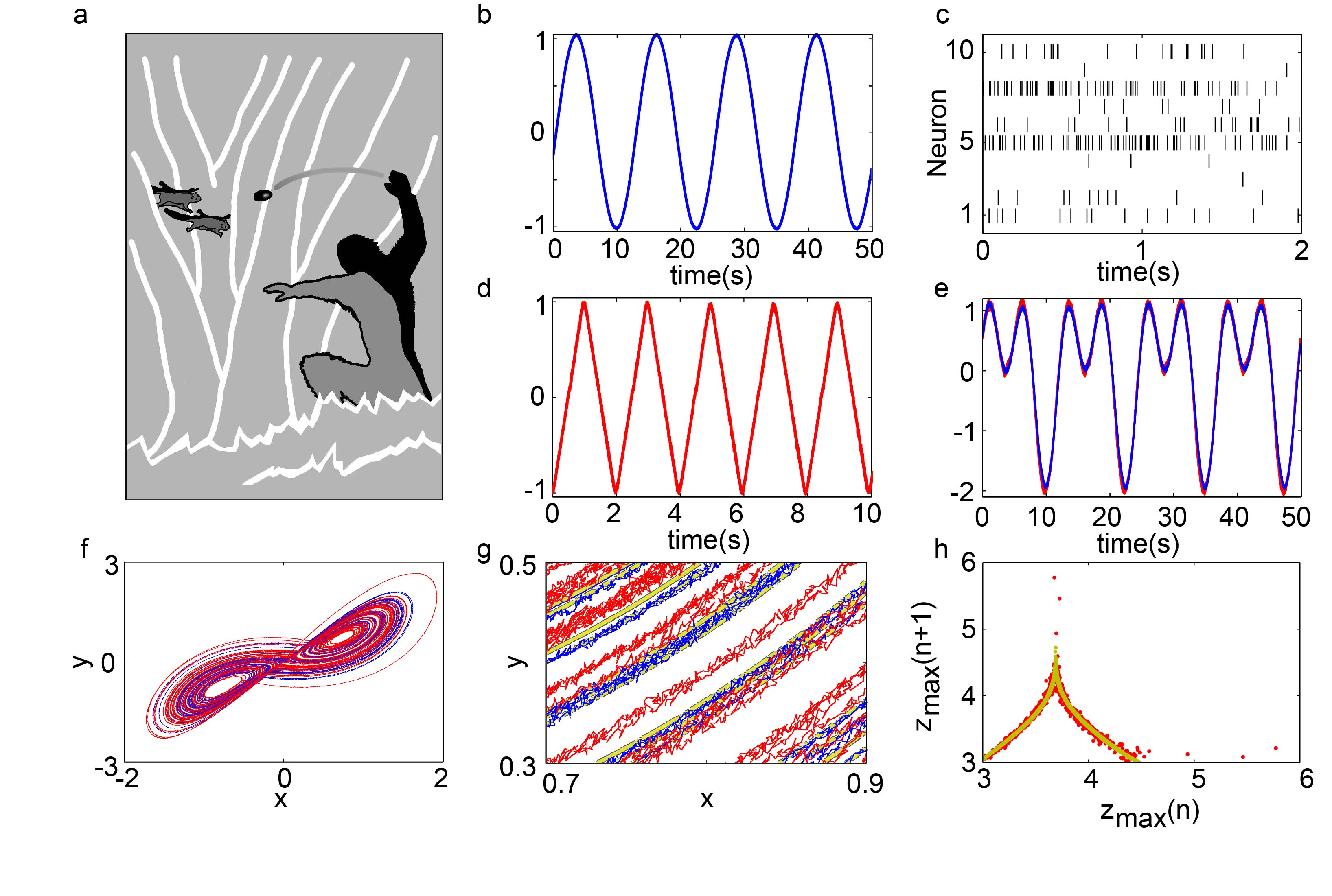}
\par\end{centering}

\protect\caption{
\textbf{Learning dynamics with spiking neural networks.} (a): Schematic
hunting scene, illustrating the need for complicated dynamical systems
learning and control. The hominid has to predict the motion of its
prey, and to predict and control the movements of its body and the
projectile. (b-h): Learning of self-sustained dynamical patterns by
spiking neural networks. (b): A sine wave generated by summed, synaptically
and dendritically filtered output spike trains of a PCSN with nonlinear
dendrites. (c): A sample of the network's spike trains generating
the sine in (b). (d): A saw tooth pattern generated by a PCSN with
saturating synapses. (e): A more complicated smooth pattern generated
by both architectures (blue: nonlinear dendrites, red: saturating
synapses). (f-h): Learning of chaotic dynamics (Lorenz system), with
a PCSN with nonlinear dendrites. (f): The spiking network imitates
an example trajectory of the Lorenz system during training (blue);
it continues generating the dynamics during testing (red). (g): Detailed
view of (f) highlighting how the example trajectory (yellow) is imitated
during training and continued during testing. (h): The spiking network
approximates not explicitly trained quantitative dynamical features,
like the tent map between subsequent maxima of the z-coordinate. The
ideal tent map (yellow) is closely approximated by the tent map generated
by the PCSN (red). The spiking network sporadically generates errors,
cf.~the larger loop in (f) and the outlier points in (h). Panel (h)
shows a ten times longer time series than (f), with three errors.
}
\end{figure}

\subsubsection*{Delayed reaction/time interval estimation}

For many tasks, e.g.~computations focusing on recent external input
and generation of self-sustained patterns, it is essential that the
memory of the involved recurrent networks is fading: If past states
cannot be forgotten, they lead to different states in response to
similar recent inputs. A readout that learns to extract computations
on recent input will then quickly reach its capacity limit. In neural
networks, fading memory originates on the one hand from the dynamics
of single neurons, e.g.~due to their finite synaptic and membrane
time constants; on the other hand it is a consequence of the neurons'
connection to a network \cite{TWG04,white2004short,goldman2009memory}.
In standard spiking neural network models, the overall fading memory
is short, of the order of hundreds of milliseconds \cite{ivry2004neural,joshi2005movement,mayor2005signal,wallace2013randomly}.
It is a matter of current debate how this can be extended by suitable
single neuron properties and topology \cite{MNM02,boerlin2013predictive,hennequin2014optimal,ostojic2014two}.
Many biological computations, e.g.~the simple understanding of a
sentence, require longer memory, on the order of seconds.

We find that CSNs without learning of recurrent connectivity or feedback
access such time scales. We illustrate this by means of a delayed
reaction/time estimation task: In the beginning of a trial, the network
receives a short input pulse. By imitation learning, the network output
learns to generate a desired delayed reaction. For this, it needs
to specifically amplify the input's dynamical trace in the recurrent
spiking activity, at a certain time interval. The desired response
is a Gaussian curve, representative for any type of delayed reaction.
The reaction can be generated several seconds after the input (Fig.~4a-c).

The quality of the reaction pattern depends on the connection strengths
within the network, specified by the spectral radius $g$ of the coupling
matrix divided by the leak of a single corresponding continuous unit
$\lambda_{x}$. Memory is kept best in an intermediate regime (Fig.~4b),
where the CSN stays active over long periods of time without overwriting
information. This has also been observed for continuous rate networks
\cite{sompolinsky1988chaos}. For too weak connections (Fig.~4a),
the CSN returns to the inactive state after short time, rendering
it impossible to retrieve input information later. If the connections
are too strong, (Fig.~4c), the CSN generates self-sustained, either
irregular asynchronous or oscillating activity, partly overwriting
information and hindering its retrieval. We observe that already the
memory in disconnected CSNs with synaptic saturation can last for
times beyond hundreds of milliseconds (cf.~ Fig.~E in S1 Supporting Information). This is a
consequence of the recovery current: If a neuron has spiked several
times in succession, the accumulated recovery current leads to further
spiking (and further recovery current), and thus dampens the decay
of a strong activation of the neuron \cite{seung2000autapse:}.

Experiments show that during time estimation tasks, neurons are particularly
active at two times: When the stimulus is received and when the estimated
time has passed \cite{deadwyler2006temporal,matell2003interval}.
Often the neuron populations that show activity at these points are
disjoint. Our model reproduces this behavior for networks with good
memory performance. In particular, at the time of the initial input
the recurrently connected neurons become highly active (gray traces
in Fig.~4b, upper sub-panel) while at the estimated reaction time,
readout neurons would show increased activity (red trace).

\begin{figure}
\noindent \begin{centering}
\includegraphics[width=1\textwidth]{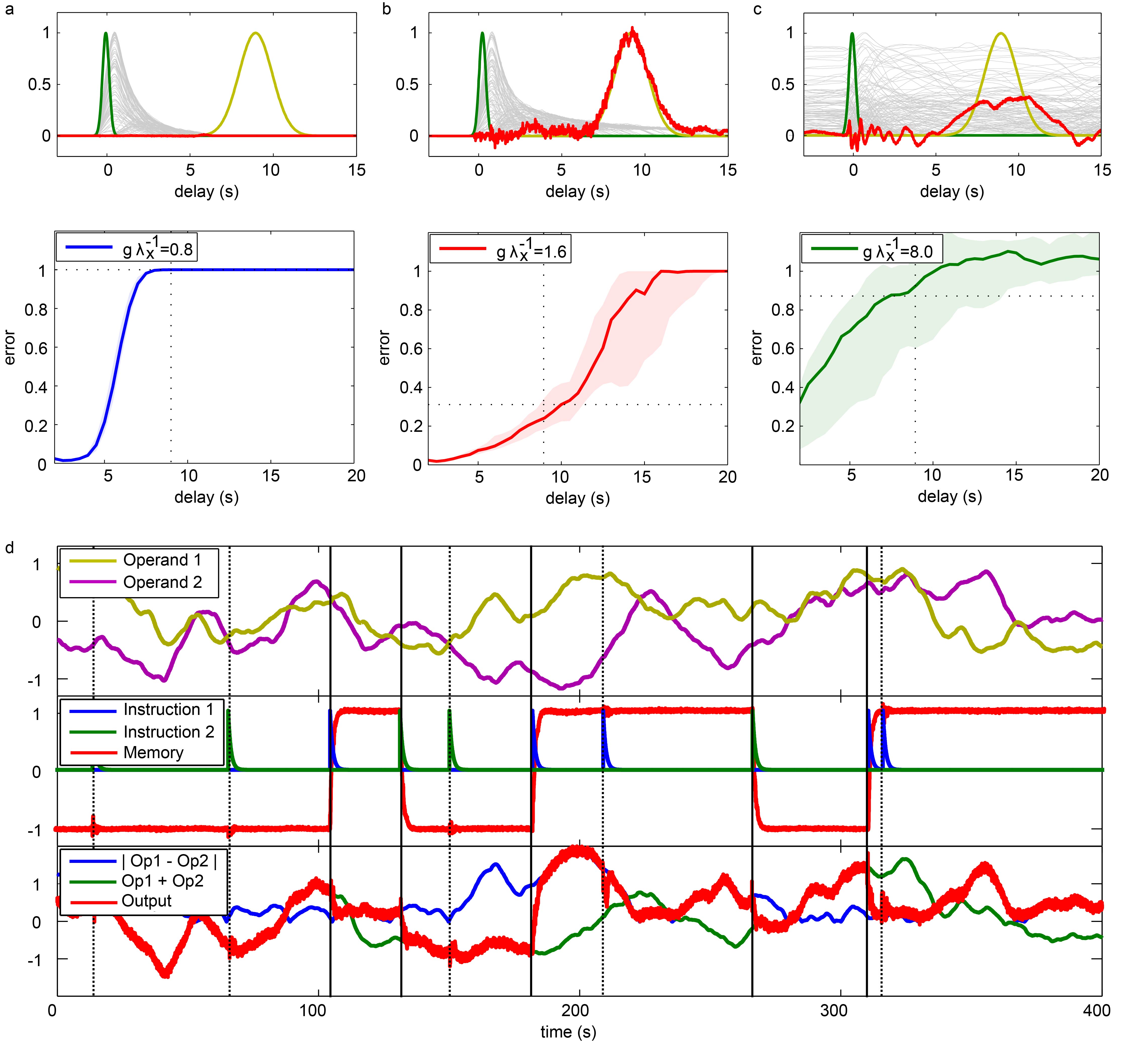}
\par\end{centering}

\protect\caption{
\textbf{Learning of longer-term memory dependent computations with
spiking neural networks.} (a-c): Delayed reaction and time interval
estimation: The synaptic output of a CSN learns to generate a generic
reaction several seconds after a short input. Upper panels show typical
examples of input, desired and actual reactions (green, yellow and
red traces). In the three panels, the desired reaction delay is the
same (9sec), the networks (CSNs with saturating synapses) have different
levels of recurrent connection strengths ((a), (b), (c): low, intermediate,
high level). The generation of the reaction is best for the network
with intermediate level of connection strength. The CSNs with lower
or higher levels have not maintained sufficient memory due to their
extinguished or noisy and likely chaotic dynamics (gray background
lines: spike rates of individual neurons). The median errors of responses
measured for different delays in ensembles of networks (levels of
connection strength as in the upper panels), are given in the lower
panels. The shaded regions represent the area between the first and
third quartile of the response errors. Dashed lines highlight delay
and error size of the examples in the upper panels. (d): Persistent
retaining of instructions and switching between computations: The
network receives (i) two random continuous operand inputs (upper sub-panel,
yellow and purple traces), and (ii) two pulsed instruction inputs
(middle sub-panel, blue and green; memory of last instruction pulse:
red). The network has learned to perform different computations on
the operand inputs, depending on the last instruction (lower subpanel):
if it was +1 (triggered by instruction channel 1), the network performs
a nonlinear computation, it outputs the absolute value of the difference
of the operands (red trace (network output) agrees with blue); if
it was -1 (triggered by channel 2), the values of the operands are
added (red trace agrees with green trace).
}
\end{figure}

\subsubsection*{Persistent memory and context dependent switching}

Tasks often also require to store memories persistently, e.g.~to
remember instructions \cite{sakai2003prefrontal}. Such memories may
be maintained in learned attractor states (e.g.~\cite{barak2014Working,H82,litwin-kumar2014Formation,zenke2015Diverse}).
In the framework of our computing scheme, this requires the presence
of output feedback \cite{maass2007computational}. Here, we illustrate
the ability of PCSNs to learn and maintain persistent memories as
attractor states as well as the ability to change behavior according
to them. For this, we use a task that requires memorizing computational
instructions (Fig.~4d) \cite{maass2007computational}. The network
has two types of inputs: After pulses in the instruction channels,
it needs to switch persistently between different rules for computation
on the current values of operand channels. To store persistent memory,
the recurrent connections are trained such that an appropriate output
can indicate the instruction channel that has sent the last pulse:
The network learns to largely ignore the signal when a pulse arrives
from the already remembered instruction channel, and to switch states
otherwise. Due to the high signal-to-noise ratio of our deterministic
spike code, the PCSNs are able to keep a very accurate representation
of the currently valid instruction in their recurrent dynamics. Fig.~4d,
middle sub-panel, shows this by displaying the output of the linear
readout trained to extract this instruction from the network dynamics.
A similarly high precision can be observed for the output of the computational
task, cf.~Fig.~4d, lower sub-panel.

\subsubsection*{Building of world models, and control}

In order to control its environment, an animal has to learn the laws
that govern the environment's dynamics, and to develop a control strategy.
Since environments are partly unpredictable and strategies are subject
to evolutionary pressure, we expect that they may be described by
stochastic optimal control theory. A particularly promising candidate
framework is path integral control, since it computes the optimal
control by simulating possible future scenarios under different random
exploratory controls, and the optimal control is a simple weighted
average of them \cite{kappen2005linear}. For this, an animal needs
an internal model of the system or tool it wants to act on. It can
then mentally simulate different ways to deal with the system and
compute an optimal one. Recent experiments indicate that animals indeed
conduct thought experiments exploring and evaluating possible future
actions and movement trajectories before performing one \cite{pfeiffer2013hippocampal,VDMDR10}.

Here we show that by imitation learning, spiking neural networks,
more precisely PCSNs with a feedback loop, can acquire an internal
model of a dynamical system and that this can be used to compute optimal
controls and actions. As a specific, representative task, we choose
to learn and control a stochastic pendulum (Fig.~5a,b). The pendulum's
dynamics are given by 
\begin{equation}
\ddot{\phi}(t)+c\omega_{0}\dot{\phi}(t)+\omega_{0}^{2}\sin(\phi(t))=\xi(t)+u(t),\label{eq:PendulumEq1}
\end{equation}
with the angular displacement $\phi$ relative to the direction of
gravitational acceleration, the undamped angular frequency for small
amplitudes $\omega_{0}$, the damping ratio $c$, a random (white
noise) angular force $\xi(t)$ and the deterministic control angular
force $u(t)$, both applied to the pivot axis. The PCSN needs to learn
the pendulum's dynamics under largely arbitrary external control forces;
this goes beyond the tasks of the previous sections. It is achieved
during an initial learning phase characterized by motor babbling as
observed in infants \cite{CVH82} and similarly in bird song learning
\cite{FFS07}: During this phase, there is no deterministic control,
$u=0$, and the pendulum is driven by a random exploratory force $\xi$
only. Also the PCSN receives $\xi$ as input and learns to imitate
the resulting pendulum's dynamics with its output.

During the subsequent control phase starting at $t=0$, the aim is
to swing the pendulum up and hold it in the inverted position (Fig.~5c).
For this, the PCSN simulates at time $t$ a set of $M$ future trajectories
of the pendulum, for different random exploratory forces $\xi_{i}$
(``mental exploration'' with $u=0$, cf.~Fig.~5a,b), starting with
the current state of the pendulum. In a biological system, the initialization
may be achieved through sensory input taking advantage of the fact
that an appropriately initialized output enslaves the network through
the feedback. Experiments indicate that explored trajectories are
evaluated, by brain regions separate from the ones storing the world
model \cite{lansink2008preferential,vandermeer2009covert,LGLMP09}.
We thus assign to the simulated trajectories a reward $R_{i}$ measuring
the agreement of the predicted states with the desired ones. The optimal
control $u(t+s)$ (cf.~Equation~\eqref{eq:PendulumEq1}) for a subsequent,
not too large time interval $s\in[0,\delta]$ is then approximately
given by a temporal average over the initial phase of the assumed
random forces, weighted by the exponentiated total expected reward,

\begin{equation}
u(t+s)=\sum_{i=1}^{M}\frac{e^{\lambda_{c}R_{i}(t)}}{\sum_{j=1}^{M}e^{\lambda_{c}R_{j}(t)}}\bar{\xi}_{i}(t),\label{eq:MainTextControlFinal}
\end{equation}
where $\bar{\xi}_{i}(t)=\frac{1}{\delta}\intop_{t}^{t+\delta}\xi_{i}(\tilde{t})d\tilde{t}$
and $\lambda_{c}$ is a weighting factor. We have chosen $R_{i}(t)=\intop_{t}^{t+T_{r}}y_{i}(\tilde{t})d\tilde{t}$,
i.e.~the expected reward increases linearly with the heights $y_{i}(\tilde{t})=-\cos(\phi_{i}(\tilde{t}))$
predicted for the pendulum for input trajectory $\xi_{i}$; it becomes
maximal for a trajectory at the inversion point. $T_{r}$ is the duration
of a simulated trajectory. The optimal control is applied to the pendulum
until $t+\Delta$, with $\Delta<\delta$. Then, at $t+\Delta$, the PCSN simulates a new set of trajectories
starting with the pendulum's updated state and a new optimal control
is computed. This is valid and applied to the pendulum between $t+\Delta$
and $t+2\Delta$, and so on. We find that controlling the pendulum
by this principle leads to the desired upswing and stabilization in
the inversion point, even though we assume that the perturbing noise
force $\xi$ (Equation~\eqref{eq:PendulumEq1}) acting on the pendulum
in addition to the deterministic control $u$, remains as strong as
it was during the exploration/learning phase (cf.~Fig.~5a,b).

\begin{figure}
\noindent \begin{centering}
\includegraphics[width=1\textwidth]{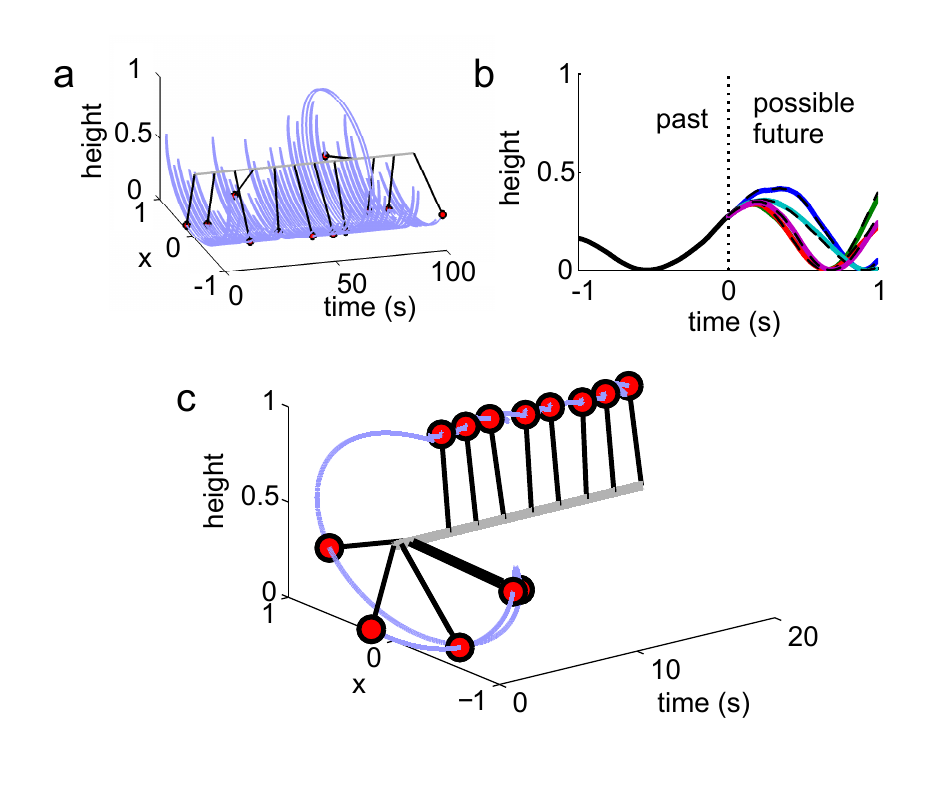}
\par\end{centering}

\protect\caption{
\textbf{Model building and mental exploration to compute optimal control.}
(a): Learning of an internal world model with spiking neural networks.
During model building, random exploratory control drives the dynamical
system (here: a swinging pendulum). The spiking neural network is
provided with the same control as input and learns to mimic the behavior
of the pendulum as its output. (b): After learning, the spiking network
can simulate the system's response to control signals. The panel displays
the height of the real pendulum in the past (solid black line) and
future heights under different exploratory controls (dashed lines).
For the same controls, the spiking neural network predicts very similar
future positions (colored lines) as the imitated system. It can therefore
be used for mental exploration and computation of optimal control
to reach an aim, here: to invert the pendulum. (c): During mental
exploration, the network simulates in regular time intervals a set
of possible future trajectories for different controls, starting from
the actual state of the pendulum. From this, the optimal control until
the next exploration can be computed and applied to the pendulum.
The control reaches its aim: The pendulum is swung up and held in
inverted position, despite a high level of noise added during testing
(uncontrolled dynamics as in panel (a)).
}
\end{figure}

We find that for controlling the pendulum, the learned internal model
of the system has to be very accurate. This implies that particular
realizations of the PCSN can be unsuited to learn the model (we observed
this for about half of the realizations), a phenomenon that has also
been reported for small continuous rate networks before. However,
we checked that continuous rate networks as encoded by our spiking
ones reliably learn the task. Since the encoding quality increases
with the number of spiking neurons, we expect that sufficiently large
PCSNs reliably learn the task as well.

\section*{Discussion}

The characteristic means of communication between neurons in the nervous
system are spikes. It is widely accepted that sequences of spikes
form the basis of neural computations in higher animals. How computations
are performed and learned is, however, largely unclear. Here we have
derived \emph{continuous signal coding spiking neural networks (CSNs),}
a class of mesoscopic spiking neural networks that are a suitable
substrate for computation. Together with plasticity rules for their
output or recurrent connections, they are able to learn general, complicated
computations by imitation learning (plastic CSNs, \emph{PCSNs}). Learning
can be highly reliable and accurate already for comparably small networks
of hundreds of neurons. The underlying principle is that the networks
reflect the input in a complicated nonlinear way, generate nonlinear
transformations of it and use fading memory such that the inputs and
their pasts interfere with each other. This requires an overall nonlinear
relaxation dynamics suitable for computations \cite{JH04}. Such dynamics
are different from standard spiking neural network dynamics, which
are characterized by a high level of noise and short intrinsic memory
\cite{joshi2005movement,mayor2005signal,JMT09,wallace2013randomly}.

To find spiking networks that generate appropriate dynamics, we use
a linear decoding scheme for continuous signals encoded in the network
dynamics as combinations of membrane potentials and synaptic currents.
A specific coding scheme like this was introduced in refs.~\cite{boerlin2011spike,boerlin2013predictive}
to derive spiking networks encoding linear dynamics in an optimal
way. We introduce spiking networks where the encoded signals have
dynamics desirable for computation, i.e.~a nonlinear, high-dimensional,
low-noise, relaxational character as well as significant fading memory.
We conclude that, since we use simple linear decoding, already the
dynamics of the spiking networks must possess these properties.

Using this approach, we study two types of CSNs: Networks with saturating
synapses and networks with nonlinear dendrites. The CSNs with saturating
synapses use a direct signal encoding; each neuron codes for one continuous
variable. It requires spiking dynamics characterized by possibly intermittent
phases of high rate spiking, or bursting, with inter-spike-intervals
smaller than the synaptic time constants, which leads to a temporal
averaging over spikes. Dynamics that appear externally similar to
such dynamics were recently highlighted as a `second type of balanced
state' in networks of pulse-coupled, intrinsically oscillating model
neurons \cite{ostojic2014two}. Very recently
\cite{harish2015asynchronous,kadmon2015transition} showed that networks
whose spiking dynamics are temporally averaged due to slow synapses
possess a phase transition from a fixed point to chaotic dynamics
in the firing rates, like the corresponding rate models that they
directly encode. In the analytical computations the spike coding was
not specified \cite{harish2015asynchronous} or assumed to be Poissonian
\cite{kadmon2015transition}. Numerical simulations of leaky integrate-and-fire
neurons in the chaotic rate regime can generate intermittent phases
of rather regular high-rate spiking \cite{harish2015asynchronous}.
The networks might provide a suitable substrate
for learning computations as well. However, since the chaotic rate
dynamics have correlations on the time scale of the slow synapses its
applicability is limited to learning tasks where only a short fading
memory of the reservoir is needed. For example delayed reaction tasks
as illustrated in Fig.~4a-c would not be possible. Interestingly,
in our scheme a standard leaky integrate-and-fire neuron with saturating
synapses appears as a special case with recovery current of amplitude
zero. According to our analysis it can act as a leaky integrator with
a leak of the same time constant as the synapses, $\lambda_{x}=\lambda_{s}$.
In contrast, in presence of a recovery current, our networks with
saturating synapses can encode slower dynamics on the order of seconds.
After training the network, the time scales can be further extended.

In the CSNs with nonlinear dendrites the entire neural population
codes for a usually smaller number of continuous variables, avoiding
high firing rates in sufficiently large networks. The networks generate
irregular, low frequency spiking and simultaneously a noise-reduced
encoding of nonlinear dynamics, the temporal
averaging over spikes in the direct coding case is partially replaced
by a spatial averaging over spike trains from many neurons. The population
coding scheme and our derivations of CSNs with nonlinear dendrites
generalize the predictive coding proposed in ref.~\cite{boerlin2013predictive}
to nonlinear dynamics. The roles of our
slow and fast connections are similar to those used there: In particular,
redundancies in the spiking are eliminated by fast recurrent connections
without synaptic filtering. We expect that these couplings can
be replaced by fast connections that have small finite synaptic time
constants, as shown for the networks of ref.~\cite{boerlin2013predictive}
in ref.~\cite{schwemmer2015constructing}. In contrast to previous
work, in the CSNs with nonlinear dendrites we have linear and nonlinear
slow couplings. The former contribute to coding precision and implement
linear parts of the encoded dynamics, the latter implement the nonlinearities
in the encoded dynamics. Further, in contrast to previous work, the
spike coding networks provide only the substrate
for learning of general dynamical systems by adapting their recurrent
connections. Importantly, this implies (i) that the neurons do not
have to adapt their nonlinearities to each nonlinear dynamical system
that is to be learned (which would not seem biologically plausible)
and (ii) that the CSNs do not have to provide a faithful approximation
of the nonlinear dynamics Equations \eqref{eq:xdot},\eqref{eq:xdotDistributed},
since a rough dynamical character (i.e. slow dynamics and the echo
state property) is sufficient for serving as substrates. We note that
refs.~\cite{eliasmith2005unified,eliasmith2012large} suggested to
use the differential equations that characterize dynamical systems
to engineer spiking neural networks that encode the dynamics. The
approach suggests an alternative derivation of spiking networks that
may be suitable as substrate for learning computations. Their rate coding
scheme, however, allows for redundancy and thus higher noise levels,
and it generates high frequency spiking. In a future publication,
B.~DePasquale, M.~Churchland, and L.F.~Abbott will present an approach
to train rate coding spiking neural networks, with continuous rate
networks providing the target signals \cite{depasquale2016using}.
We will discuss the relation between our and this approach in a joint
review \cite{abbott2016building}.

A characteristic feature of our neuron models is that they take into
account nonlinearities in the synapses or in the dendrites. On the one
hand this is biologically plausible
\cite{blitz2004short,LH05,Memmesheimer2010,caze2013passive}, on the
other hand it is important for generating nonlinear computations.  Our
nonlinearities are such that the decoded continuous dynamics match
those for typical networks of continuous rate neurons and provide a
simple model for dendritic and synaptic saturation. However, the
precise form of the neuron model and its nonlinearity is not important
for our approaches: As long as the encoded dynamical system is
suitable as a computational reservoir, the spiking system is a CSN and
our learning schemes will work. As an example, a dendritic tree with
multiple interacting compartments may be directly implemented in
both the networks with saturating synapses and in the networks with
nonlinear dendrites. A future task is to explore the computational
capabilities of CSNs incorporating different and biologically more
detailed features that lead to nonlinearities, e.g.~neural refractory
periods, dendritic trees with calcium and NMDA voltage dependent
channels and/or standard types of short term synaptic plasticity.

Inspired by animals' needs to generate and predict continuous dynamics
such as their own body and external world movements, we let our networks
learn to approximate desired continuous dynamics. Since effector organs
such as muscles and post-synaptic neurons react to weighted, possibly
dendritically processed sums of post-synaptic currents, we interpret
these sums as the relevant, continuous signal-approximating outputs
of the network \cite{eliasmith2003neural}. Importantly, this is not
the same as Poissonian rate coding of a continuous signal: As a simple
example, consider a single spiking neuron. In our scheme it will spike
with constant inter-spike-intervals to encode a constant output. In
Poissonian rate coding, the inter-spike-intervals will be random,
exponentially distributed and many more spikes need to be sampled
to decode the constant output (cf.~Fig.~A in S1 Supporting Information).

The outputs and recurrent connections of CSNs can be learned by standard
learning rules \cite{HS02,SA09}. The weight changes depend on the
product of the error and the synaptic or dendritic currents and may
be interpreted as delta-rules with synapse- and time-dependent learning
rates. PCSNs, with learning of recurrent weights or output feedback,
show how spiking neural networks may learn internal models of complicated,
self-sustained environmental dynamics. In our applications, we demonstrate
that they can learn to generate and predict the dynamics in different
depths, ranging from the learning of single stable patterns over the
learning of chaotic dynamics to the learning of dynamics incorporating
their reactions to external influences.

The spiking networks we use have medium size, like networks with continuous
neurons used in the literature \cite{JH04,SA09}. CSNs with saturating
synapses have, by construction, the same size as their non-spiking
counterparts. In CSNs with nonlinear dendrites the spike load necessary
to encode the continuous signals is distributed over the entire network.
This leads to a trade-off between lower spiking frequency per neuron
and larger network size (cf. Fig.~F in S1 Supporting Information): The
faster the neurons can spike the smaller the network may be to solve
a given task.

Previous work using spiking neurons as a reservoir to generate a high
dimensional, nonlinear projection of a signal for computation, concentrated
on networks without output feedback or equivalent task-specific learning
of recurrent connectivity \cite{MNM02,LM07,hennequin2014optimal}.
Such networks are commonly called ``liquid state machines'' \cite{maass2010liquid}.
By construction, they are unable to solve tasks like the generation
of self-sustained activity and persistent memorizing of instructions;
these require an effective output feedback, since the current output
determines the desired future one: To compute the latter, the former
must be made available to the network as an input. The implementation
of spiking reservoir computers with feedback was hindered by the high
level of noise in the relevant signals: The computations depend on
the spike rate, the spike trains provide a too noisy approximation
of this average signal and the noise is amplified in the feedback
loop. While analytically considering feedback in networks of continuous
rate neurons, ref.~\cite{maass2007computational} showed examples
of input-output tasks solved by spiking networks with a feedback circuit,
the output signals are affected by a high level of noise. This concerns
even output signals just keeping a constant value. We implemented
similar tasks (Fig.~4d), and find that our networks solve them very
accurately due to their more efficient coding and the resulting comparably
high signal-to-noise ratio. In contrast to previous work, our derivations
systematically delineate spiking networks which are suitable for the
computational principle with feedback or recurrent learning; the networks
can accurately learn universal, complicated memory dependent computations
as well as dynamical systems approximation, in particular the generation
of self-sustained dynamics.

In the control task, we show how a spiking neural network can learn
an internal model of a dynamical system, which subsequently allows
to control the system. We use a path integral approach, which has
already previously been suggested as a theory for motor control in
biological systems \cite{Friston2011488,Todorov14072009}. We apply
it to learned world models, and to neural networks. Path integral
control assumes that noise and control act in a similar way on the
system \cite{kappen2005linear}. This assumption is comparably weak
and the path integral control method has been successfully applied
in many robotics applications \cite{Theodorou_RAIIC_2010,Pastor_RAIIC_2011,Buchli-RSS-10},
where it was found to be superior to reinforcement learning and adaptive
control methods.

Continuous rate networks using recurrence, readouts, and feedback
or equivalent recurrent learning, are versatile, powerful devices
for nonlinear computations. This has inspired their use in manifold
applications in science and engineering, such as control, forecasting
and pattern recognition \cite{lukosevicius2012reservoir}. Our study
has demonstrated that it is possible to obtain similar performance
using spiking neural networks. Therewith, our study makes spiking
neural networks available for similarly diverse, complex computations
and supports the feasibility of the considered computational principle as a principle
for information processing in the brain.

\section*{Methods}

\subsection*{Network simulation}

We use a time grid based simulation scheme (step size $dt$). If not
mentioned otherwise, between time points, we compute the membrane
potentials using a Runge-Kutta integration scheme for dynamics without
noise and an Euler-Maruyama integration scheme for dynamics with noise.
Since CSNs with nonlinear dendrites have fast connections without
conduction delays and synaptic filtering, we process spikings at a
time point as follows: We test whether the neuron with the highest
membrane potential is above threshold. If the outcome is positive,
the neuron is reset and the impact of the spike on postsynaptic neurons
is evaluated. Thereafter, we compute the neuron with the highest,
possibly updated, membrane potential and repeat the procedure. If
all neurons have subthreshold membrane potential, we proceed to the
next time point. The described consecutive updating of neurons in
a single time step increases in networks with nonlinear dendrites
the robustness of the simulations against larger time steps, as the
neurons maintain an order of spiking and responding like in a simulation
with smaller time steps and a small but finite conduction delay and/or
slight filtering of fast inputs. As an example, the scheme avoids
that neurons that code for similar features and thus possess fast
mutual inhibition, spike together within one step and generate an
overshoot in the readout, as it would be the case in a parallel membrane
potential updating scheme. The different tasks use either networks
with saturating synapses or networks with nonlinear dendrites. In
both cases, $\mathbf{A}$ is a sparse matrix with a fraction $p$
of non-zero values. These are drawn independently from a Gaussian
distribution with zero mean and variance $\frac{g^{2}}{pN}$ (CSNs
with saturating synapses) or $\frac{g^{2}}{pJ}$ (CSNs with nonlinear
dendrites), which sets the spectral radius of $\mathbf{A}$ approximately
to $g$. For networks with nonlinear dendrites, the elements of $\mathbf{\boldsymbol{\Gamma}}$
are drawn from a standard normal distribution. To keep the approach
simple, we allow for positive and negative dendro-somatic couplings.
In order to achieve a uniform distribution of spiking over the neurons
in the network, we normalize the columns of $\mathbf{\mathbf{\boldsymbol{\Gamma}}}$
to have the same norm, which we control with the parameter $\gamma_{s}$. This
implies that the thresholds are identical.

\subsection*{Training phase}

The networks are trained for a period of length $T_{t}$ such that
the readouts $z_{k}$ imitate target signals $F_{k}(t)$, i.e.~such
that the time average of the square of the errors $e_{k}(t)=z_{k}(t)-F_{k}(t)$
is minimized. At $T_{t}$, training stops and the weights are not
updated anymore in the subsequent testing. If present, the external
input to the neurons is a weighted sum of $K_{\text{in}}$ continuous
input signals $f_{k}(t)$, $I_{\text{e,}\beta}(t)=\sum_{k=1}^{K_{\text{in}}}w_{\beta k}f_{k}(t)$,
where the index $\beta$ runs from $1$ to $N$ (CSNs with saturating
synapses) or from $1$ to $J$ (CSNs with nonlinear dendrites). The
weights $w_{\beta k}$ are fixed and drawn from a uniform distribution
in the range $[-\tilde{w}^{i},\tilde{w}^{i}]$. If present, the feedback
weights $w_{\beta k}^{f}$ (cf.~Equation~\eqref{eq:feedback-1-1})
are likewise chosen randomly from a uniform distribution in the range
$[-\widetilde{w}^{f},\widetilde{w}^{f}]$ with a global feedback parameter
$\widetilde{w}^{f}$.

For the delayed reaction/time estimation task (Figs.~4a-c, E in S1 Supporting Information), we
applied the RLS (recursive least squares) algorithm \cite{HS02} to
learn the linear outputs. For the pattern generation, instruction
switching and control tasks, we applied the FORCE (first-order reduced
and controlled error) algorithm \cite{SA09} (Figs.~3, 4d, 5, A-D, F and G in S1 Supporting Information) to learn the recurrent connections and linear outputs.

\subsection*{Learning rules}

The output weights $w_{km}^{o}$ are trained using the standard recursive
least squares method \cite{HS02}. They are initialized with $0$,
we use weight update intervals of $\Delta t$. The weight update uses
the current training error $e_{k}(t)=z_{k}(t)-F_{k}(t)$, where $z_{k}(t)$
is the output that should imitate the target signal $F_{k}(t)$, it
further uses an estimate $P_{\beta\rho}(t)$ of the inverse correlation
matrix of the unweighted neural synaptic or dendritic inputs $\tilde{r}_{\beta}(t)$,
as well as these inputs, 
\begin{equation}
w_{k\beta}^{o}(t)=w_{k\beta}^{o}(t-\Delta t)-e_{k}(t)\sum_{\rho}P_{\beta\rho}(t)\widetilde{r}_{\rho}(t).\label{eq:wmk-1}
\end{equation}

The indices $\beta,\rho$ range over all saturating synapses ($\beta,\rho=1,...,N$;
$\tilde{r}_{\beta}(t)=\tanh\left(\gamma r_{\beta}(t)\right)$) or
all non-linear dendrites ($\beta,\rho=1,...,J$; $\tilde{r}_{\beta}(t)=\tanh\left(\sum_{m=1}^{N}\Gamma_{\beta m}r_{m}(t)\right)$)
of the output neuron. The square matrix $P$ is a running filter estimate
of the inverse correlation matrix of the activity of the saturated
synapses (CSNs with saturating synapses) or non-linear dendrites (CSNs
with nonlinear dendrites). The matrix is updated via 
\begin{equation}
P_{\beta\gamma}(t)=P_{\beta\gamma}(t-\Delta t)-\frac{\sum_{\rho}\sum_{\sigma}P_{\beta\rho}(t-\Delta t)\widetilde{r}_{\rho}(t)\widetilde{r}_{\sigma}(t)P_{\sigma\gamma}(t-\Delta t)}{1+\sum_{\rho}\sum_{\sigma}\widetilde{r}_{\rho}(t)P_{\rho\sigma}(t-\Delta t)\widetilde{r}_{\sigma}(t)},\label{eq:Pin-1}
\end{equation}
where the indices $\beta,\gamma,\rho,\sigma$ run from $1$ to $N$
(CSNs with saturating synapses) or from $1$ to $J$ (CSNs with nonlinear
dendrites). $\mathbf{P}$ is initialized as $\mathbf{P}(0)=\alpha^{-1}\mathbf{1}$
with $\alpha^{-1}$ acting as a learning rate.

For the update of output weights in presence of feedback and of recurrent
weights we adopt the FORCE algorithm \cite{SA09}. In presence of
feedback, this means that recursive least squares learning of output
is fast against the temporal evolution of the network, and already
during training the output is fed back into the network. Thus, each
neuron gets a feedback input 
\begin{equation}
I_{\text{e},\beta}^{f}(t)=\sum_{k=1}^{K_{\text{out}}}w_{\beta k}^{f}z_{k}(t)=\sum_{k=1}^{K_{\text{out}}}w_{\beta k}^{f}\sum_{\rho}w_{k\rho}^{o}\widetilde{r}_{\rho}(t).\label{eq:feedback-1}
\end{equation}
The feedback weights $w_{\beta k}^{f}$ are static, the output weights
are learned according to Equation~\eqref{eq:wmk-1}.

Since the outputs are linear combinations of synaptic or dendritic
currents, which also the neurons within the network linearly combine,
the feedback loop can be implemented by modifying the recurrent connectivity,
by adding a term $\sum_{k=1}^{K_{\text{out}}}w_{\rho k}^{f}w_{k\beta}^{o}$
to the matrix $A_{\rho\beta}$. Learning then affects the output weights
as well as the recurrent connections, separate feedback connections
are not present. This learning and learning of output weights with
a feedback loop are just two different interpretations of the same
learning rule. For networks with saturating synapses the update is
\begin{equation}
A_{nm}(t)=A_{nm}(t-\Delta t)-\sum_{k=1}^{K_{\text{out}}}w_{nk}^{f}e_{k}(t)\sum_{l=1}^{N}P_{ml}(t)\widetilde{r}_{l}(t),\label{eq:updateA-1}
\end{equation}
where the $w_{nk}^{f}$ are now acting
as learning rates. For networks with nonlinear dendrites, the update
is 
\begin{equation}
D_{nj}(t)=D_{nj}(t-\Delta t)-\sum_{i=1}^{J}\Gamma_{in}\sum_{k=1}^{K_{\text{out}}}w_{ik}^{f}e_{k}(t)\sum_{h=1}^{J}P_{jh}(t)\widetilde{r}_{h}(t).\label{eq:Djm-1}
\end{equation}

\subsection*{Control task}

The task is achieved in two phases, the learning and the control phase.

1.~Learning: The PCSN learns a world model of the noisy pendulum,
i.e.~it learns the dynamical system and how it reacts to input. The
pendulum follows the differential Equation \eqref{eq:PendulumEq1}
with $c\omega_{0}=0.1\text{s}^{-1}$ and $\omega_{0}^{2}=10\text{s}^{-2}$,
$\xi(t)$ is a white noise force with $\left\langle \xi(t)\xi(t')\right\rangle =\text{s}^{-3}\delta(t-t')$,
$x(t)=\sin(\phi(t))$ and $y(t)=-\cos(\phi(t))$
are Cartesian coordinates of the point mass. The neural network has
one input and three outputs which are fed back into the network; it
learns to output the $x$- and the $y$-coordinate, as well as the
angular velocity of the pendulum when it receives as input the strength
of the angular force (noise plus control) $\xi(t)+u(t)$ applied to
the pivot axis of the pendulum. The learning is here interpreted as
learning in a network with feedback, cf.~Equation~\eqref{eq:feedback-1}.

We created a training trajectory of length $T_{t}=1000$s by simulating
the pendulum with the given parameters and by driving it with white
noise $\xi(t)$ as an exploratory control ($u(t)=0$). Through its
input, the PCSN receives the same white noise realization $\xi(t)$.
During training the PCSN learns to imitate the reaction of the pendulum
to this control, more precisely its outputs learn to approximate the
trajectories of $x$, $y$ and $\omega$. As feedback to the reservoir
during training we choose a convex combination of the reservoir output
and the target ($\text{\text{feedback}}=0.9\cdot\text{output}+0.1\cdot\text{target}$).
We find that such a combination improves performance: If the output
at the beginning of the training is very erroneous, those errors are
accumulated through the feedback-loop, which prevents the algorithm
from working. On the other hand, if one feeds back only the target
signal, the algorithm does not learn how to correct for feedback transmitted
readout errors. In our task, the convex combination alleviates both
problems.

2.~Control: In the second phase, the learned world model of the pendulum
is used to compute stochastic optimal control that swings the pendulum
up and keeps it in the inverted position. The PCSN does not learn
its weights in this phase anymore. It receives the different realizations
of exploratory (white noise) control and predicts the resulting motion
(``mental exploration''). From this, the optimal control may be
computed using the path integral framework \cite{kappen2005linear}.
In this framework a stochastic dynamical system (which is possibly
multivariate) 
\begin{align}
\dot{\mathbf{x}}(t)= & \mathbf{f}(\mathbf{x}(t))+\mathbf{u}(\mathbf{x}(t),t)+\mathbf{\boldsymbol{\xi}}(t)\label{eq:xdotPI}
\end{align}
with arbitrary nonlinearity $\mathbf{f}(\mathbf{x}(t))$ and white
noise $\mathbf{\boldsymbol{\xi}}(t)$, is controlled by the feedback
controller $\mathbf{u}(\mathbf{x}(t),t)$ to optimize an integral
$C(t)$ over a state cost $U(x(\tilde{t}))$ and a moving horizon
quadratic control cost, $C(t)=\intop_{t}^{t+T_{r}}U(\mathbf{x}(\tilde{t}))+\mathbf{u}(\tilde{t})^{2}d\tilde{t}$.
The reward is related to the cost by $R=-C$. Path integral control
theory shows that the control at time $t$ can be computed by generating
samples from the dynamical system under the uncontrolled dynamics
\begin{align}
\dot{\mathbf{x}}(t)= & \mathbf{f}(\mathbf{x}(t))+\mathbf{\boldsymbol{\xi}}(t).\label{eq:xdotuncontrolledPI-1}
\end{align}
The control is then given by the success weighted average of the noise
realizations $\xi_{i}$ 
\begin{equation}
u(t)=\lim_{\delta\rightarrow0}{\lim_{M\rightarrow\infty}{\sum_{i=1}^{M}\frac{e^{-\lambda_{c}C_{i}(t)}}{\sum_{j=1}^{M}e^{-\lambda_{c}C_{j}(t)}}\frac{1}{\delta}\intop_{t}^{t+\delta}\xi_{i}(\tilde{t})d\tilde{t}}},\label{eq:-maintextcontrol-1}
\end{equation}
where $C_{i}(t)=\intop_{t}^{t+T_{r}}U(\mathbf{x}_{i}(\tilde{t}))d\tilde{t}$
is the cost observed in the $i$th realization of the uncontrolled
dynamics, which is driven by noise realization $\xi_{i}$ and $u=0$.
Equation~\eqref{eq:MainTextControlFinal} is a discrete approximation
to Equation~\eqref{eq:-maintextcontrol-1}. In our task, Equation~\eqref{eq:xdotuncontrolledPI-1}
becomes 
\begin{align*}
\dot{\phi}(t) & =\omega(t)\\
\dot{\omega}(t) & =-\omega_{0}^{2}\sin(\phi(t))-c\omega_{0}\omega(t)+\xi(t)+u(t)
\end{align*}
and $U(\mathbf{x}(t))=-y(t)=\cos(\phi(t))$.

\subsection*{Figure details}

\begin{table}
\begin{tabular}{|l||c|c|c|c|c|c|c|c|}
\hline 
Sat. syn.  & N  & $\alpha$  & dt  & $T_{t}$  & $\lambda_{s}^{-1}$  & $\lambda_{V}^{-1}$  & $\text{V}_{r}$  & $\theta$\tabularnewline
\hline 
\hline 
Fig.~3d  & $50$  & $0.1$  & $0.1$ms  & $100$s  & $100$ms  & $100$ms  & $0.9\theta$  & $0.03$\tabularnewline
\hline 
Fig.~3e  & $50$  & $0.1$  & $1$ms  & $100$s  & $100$ms  & $100$ms  & $0.9\theta$  & $0.03$\tabularnewline
\hline 
Fig.~4a-c  & $200$  & $0.1$  & $1$ms  & $800$s  & $100$ms  & $50$ms  & $0.54\theta$  & $0.1$\tabularnewline
\hline 
\end{tabular}

\protect\caption{Parameters used in the different figures for simulations of networks
with saturating synapses.\label{tab:ParametersSatSyn}}
\end{table}

\begin{table}
\begin{tabular}{|l||c|c|c|c|c|c|c|c|c|c|}
\hline 
Nonlin. dendr.  & N  & J  & $\alpha$  & $\gamma_{s}$  & dt  & $T_{t}$  & $\mu$  & $\lambda_{s}^{-1}$  & $\lambda_{V}^{-1}$  & $\text{a}$\tabularnewline
\hline 
\hline 
Fig.~3b,c  & $500$  & $50$  & $0.1$  & $0.03$  & $1$ms  & $100$s  & $0$  & $100$ms  & $100$ms  & $\lambda_{s}-1$s\tabularnewline
\hline 
Fig.~3e  & $500$  & $50$  & $0.1$  & $0.03$  & $1$ms  & $100$s  & $0$  & $100$ms  & $100$ms  & $\lambda_{s}$$-1$s\tabularnewline
\hline 
Fig.~3f-h  & $1600$  & $800$  & $0.1$  & $0.03$  & $1$ms  & $200$s  & $0$  & $100$ms  & $100$ms  & $\lambda_{s}$$-1$s\tabularnewline
\hline 
Fig.~4d  & $300$  & $300$  & $50$  & $0.5$  & $10$ms  & $1000$s  & $0$  & $1$s  & $0.5$s  & $\lambda_{s}-$$0.02$s\tabularnewline
\hline 
Fig.~5  & $500$  & $300$  & $0.1$  & $0.03$  & $1$ms  & $1000$s  & $20/N^{2}$  & $100$ms  & $50$ms  & $\lambda_{s}-$$10$s\tabularnewline
\hline 
\end{tabular}

\protect\caption{Parameters used in the different figures for simulations of networks
with nonlinear dendrites. The parameter $a=\lambda_{s}-\lambda_{x}$
is given in terms of $\lambda_{s}$ and $\lambda_{x}$\label{tab:ParametersNonDen}}
\end{table}

The parameters of the different simulations are given in Table \ref{tab:ParametersSatSyn}
for simulations using saturating synapses and in Table \ref{tab:ParametersNonDen}
for simulations using nonlinear dendrites. Further parameters and
details about the figures and simulations are given in the following
paragraphs.

If not mentioned otherwise, for all simulations we use $g=1.5\frac{1}{\text{s}}$,
$p=0.1$, $\tilde{w}^{f}=1\frac{1}{\s}$, $\tilde{w}^{i}=1\frac{1}{\s}$,
$\Delta t=0.01\text{s}$, $\gamma=\theta$ and $\sigma_{\eta}=0\frac{1}{\sqrt{\s}}$.
We note that for simulations with saturating synapses, we model the
slow synaptic currents to possess synaptic time constants of $100$ms
(cf., e.g., \cite{hennequin2014optimal,zenke2015Diverse}). We usually
use the same value for the slow synapses in networks with nonlinear
dendrites. Upon rescaling time, these networks can be interpreted
as networks with faster time constants, which learn faster target
dynamics. Since the spike rates scale likewise, we have to consider
larger networks to generate rates in the biologically plausible range
(cf.~Fig.~F in S1 Supporting Information).

\subsubsection*{Figure 3}

Figure 3b,c: The PCSN has non-linear dendrites. The target signal is
a sine with period $4\pi\text{s}$ and amplitude $2$ (normalized
to one in the figure). During recall, the neurons of the PCSN spike
with mean rate $30.2$Hz.

Figure 3d: The PCSN has saturating synapses. The target signal is
a saw tooth pattern with period $2\text{s}$ and amplitude $10$ (normalized
to one in the figure). We used an Euler scheme here. The mean spike
rate is $226$Hz.

Figure 3e: The task is performed by a PCSN with non-linear dendrites
and by a PCSN with saturating synapses. The target signal is $\sin(t\frac{0.5}{\text{s}})+\cos(t\frac{1}{\text{s}})$.
The mean spike rate is $77.8$Hz for saturating synapses and $21.3$Hz
for non-linear dendrites.

Figure 3f-h: The PCSN has nonlinear dendrites. As teacher we use the
standard Lorenz system 
\begin{align*}
\dot{x}(t) & =\sigma(y(t)-x(t))\\
\dot{y}(t) & =x(t)(\rho-z(t))-y(t)\\
\dot{z}(t) & =x(t)y(t)-\beta z(t)
\end{align*}
with parameters $\sigma=10$, $\rho=28$, $\beta=8/3$; we set the
dimensionless temporal unit to $0.2\text{s}$ and scale the dynamical
variables by a factor of $0.1$. Panels (f,g) show a recall phase
of $400$s, panel (h) shows points from a simulation of $4000$s.
Panel (f) only shows every 10th data point, panel (g) shows every
data point. The mean spike rate is $432$Hz.

\subsubsection*{Figure 4}

Figure 4a-c: We quantified the memory capacity of a CSN with saturating
synapses. The network has a sparse connectivity matrix $\mathbf{A}$
without autapses. We applied white noise with $\sigma_{\eta}=0.001\frac{1}{\sqrt{\s}}$.
The input is a Gaussian bell curve with $\sigma=0.2$s and integral
$10$s (height normalized to one in the figure). The target is a Gaussian
bell curve with $\sigma=1$s and integral $1$s (height normalized
to one in the figure). The target is presented several seconds after
the input. Trials consisting of inputs and subsequent desired outputs
are generated at random times with exponential inter-trial-interval
distribution with time constant $10$s and a refractory time of $100$s.
Training time is $T_{t}=800\text{s}$, i.e.~the network is trained
with about $6$ to $8$ trials. Testing has the same duration with
a similar number of trials. There is no feedback introduced by initialization
or by learning, so the memory effect is purely inherent to the random
network. We compute the quality of the desired output generation as
the root mean squared (RMS) error between the generated and the desired
response, normalized by the number of test trials. As reference, we
set the error of the ``extinguished'' network, which does not generate
any reaction to the input, to $1$. Lower panels of Fig.~4a-c display
medians and quartiles taken over $50$ task repetitions. The sweep was
done for time-delays $2-20$s in steps of $0.5$ s.

Figure 4d: The PCSN has nonlinear dendrites. For this task a constant
input of ${\bf I_{e}^{\text{const}}}=\mathbf{b}$ is added to the
network with the elements of the vector $\mathbf{b}$ chosen uniformly
from $[0\frac{1}{\s},250\frac{1}{\s}]$ to introduce inhomogeneity.
Four different inputs are fed into the network, two continuous $f_{1/2}^{c}$
and two pulsed input channels $f_{1/2}^{p}$. The continuous inputs
are created by convolving white noise twice with an exponential kernel
$e^{-t\frac{1}{\s}}$ (equivalent to convolving once with an alpha
function) during training and $e^{-t\frac{1}{10\s}}$ during testing.
The continuous input signals are normalized to have mean $0$ and
standard deviation $0.5$. The pulsed instruction input is created
by the convolution of a Poisson spike train with an exponential kernel
$e^{-t\frac{1}{\s}}$. The rate of the delta pulses during training
is $0.04\frac{1}{\s}$. During testing we choose a slower rate of
$0.01\frac{1}{\s}$ for a clearer presentation. In the rare case when
two pulses overlap such that the pulsed signal exceeds an absolute
value of $1.01$ times the maximal pulse height of one, we shift the
pulse by the minimal required amount of time to achieve a sum of the
pulses below or equal to $1.01$. We use weights $\widetilde{w}^{i,p}=100\frac{1}{\s}$
for the pulsed inputs, $\widetilde{w}^{i,c}=250\frac{1}{\s}$ for
the continuous inputs and $\widetilde{w}^{f}=250\frac{1}{\s}$ for
the feedback; $g=75\frac{1}{\s}$. The recurrent weights of the network
are trained with respect to the memory target $F_{m}(t)$. This target
is $+1$ if the last instruction pulse came from $f_{1}^{p}$ and
it is $-1$ if the last pulse came from $f_{2}^{p}$. During switching
the target follows the integral of the input pulse. The corresponding
readout is $z_{m}$. The second readout $z_{c}$ is trained to output
the absolute value of the difference of the two continuous inputs,
if the last instruction pulse came from $f_{1}^{p}$, and to output
their sum, if the last instruction pulse came from $f_{2}^{p}$. The
specific analytical form of this target is $F_{c}(t)=\left|f_{1}^{c}(t)-f_{2}^{c}(t)\right|(F_{m}(t)+1)/2-(f_{1}^{c}(t)+f_{2}^{c}(t))(F_{m}(t)-1)/2$.
The mean spike rate is $5.53$Hz.

\subsubsection*{Figure 5}

Since we have white noise as input we use the Euler-Maruyama scheme
in all differential equations. The PCSN has nonlinear dendrites. Non-plastic
coupling strengths are $\widetilde{w}^{f,y}=100\frac{1}{\s}$ for
the feedback of the y-coordinate, $\widetilde{w}^{f,x}=100\frac{1}{\s}$
for the feedback of the x-coordinate, $\widetilde{w}^{f,\omega}=20\frac{1}{\s}$
for the feedback of the angular velocity and $\widetilde{w}^{i}=\frac{2}{7}\frac{1}{\s}$
for the input. We introduce an additional random constant bias term
into the nonlinearity to increase inhomogeneity between the neurons:
The nonlinearity is $\tanh\left(b_{j}+\sum_{m=1}^{N}W_{njm}r_{m}(t)\right)$
where $b_{j}$ is drawn from a Gaussian distribution with standard
deviation $0.01$. The integration time $\delta$ is $0.1$s. During
the control/testing phase, every $\Delta=0.01\text{s}$, $M=200$
samples of length $T_{r}=1$s are created, the cost function is weighted
with $\lambda_{c}=0.01\frac{1}{\s}$. The mean spike rate is $146$Hz.


\section*{Acknowledgements}

We thank Marije ter Wal, Hans Ruiz, Sep Thijssen, Joris Bierkens, Mario Mulansky, Vicen\c c Gomez and
Kevin Sharp for fruitful discussions and Hans G\"unter Mem\-mes\-hei\-mer and Verena Thal\-meier for help with the graphical illustrations.

%
%

\setcounter{page}{1}
\setcounter{equation}{0}
\setcounter{table}{0}
\setcounter{figure}{0}
\renewcommand{\theequation}{S\arabic{equation}}
\renewcommand{\thetable}{\Alph{table}}
\renewcommand{\thefigure}{\Alph{figure}}

\part*{S1 Supporting information}

\newpage{}

\section{Supporting text and figures}

\subsection{Networks with nonlinear dendrites}

As stated in the main text, we can generalize Equation (3) by introducing
fast connections that generate discontinuities in postsynaptic neurons
when a neuron spikes. We may then require that only a lower, say $J-$,
dimensional combination \foreignlanguage{english}{$\mathbf{x}(t)$
of the $N-$dimensional vectors $\mathbf{V}(t)$ and $\mathbf{r}(t)$
is continuous,} 
\begin{equation}
\mathbf{x}(t)=\mathbf{L}\mathbf{V}(t)+\mathbf{\tilde{\Gamma}}\mathbf{r}(t),\label{eq:xLvGx}
\end{equation}
where $\mathbf{L}$ and $\mathbf{\tilde{\mathbf{\Gamma}}}$ are $J\times N$
matrices. (For clarity of presentation we will use vector/matrix notation
instead of components throughout the present section.) The benefit
of this approach is that the spike trains of a larger population of
neurons contribute to each $x_{n}$, such that a modified analogue
to Equation (9),
\begin{equation}
\mathbf{x}(t)\approx\mathbf{\Gamma}\mathbf{r}(t),\label{eq:xGammar}
\end{equation}
with a $J\times N$ matrix $\mathbf{\Gamma}$ can hold even if the
spike rates of individual neurons are low, i.e.~if we make use of
population/distributed coding. The matrix $\mathbf{L}$ is fixed (except
for degenerate cases) as soon as the matrix $\mathbf{\tilde{\Gamma}}$
and the fast changes in $\mathbf{V}$ are fixed. However, it is not
a priori clear how to choose the latter two; we need to employ some
optimization scheme to ensure both a good approximation Equation \eqref{eq:xGammar}
and a low firing rate.

For this, we start anew, and in contrast to the previous section with
the dynamics for $\mathbf{x}(t)$. From these we will derive spiking
dynamics approximating the $\mathbf{x}(t)$. We begin with a general
$J$-dimensional nonlinear dynamical system yielding $\mathbf{x}(t)$,
\begin{equation}
\dot{\mathbf{x}}(t)=\mathbf{f}(\mathbf{x}(t))+\mathbf{c}(t),\label{eq:xdotfxc}
\end{equation}
where $\mathbf{f}(\mathbf{x})$ and $\mathbf{c}(t)$ are column vectors
of functions $f_{j}(x_{1},...,x_{N})$ and external inputs $c_{j}(t)$,
respectively. We will generalize an approach introduced in refs.~\citep{boerlin2011spike,bourdoukan2012learning,boerlin2013predictive}
to nonlinear systems and derive spiking dynamics that optimally (see
below) approximate $\mathbf{x}(t)$ satisfying Equation \eqref{eq:xdotfxc}.
The approach will yield Equation \eqref{eq:xLvGx} with a specific
$\mathbf{L}$ as by-product. We will find that the dynamics of individual
neurons depend on the $f_{j}$ and we will specify these functions
such that the neural dynamics are biologically plausible and suitable
for universal computation.

We choose the momentary error or cost function
\begin{equation}
E(t)=\left(\mathbf{x}(t)-\mathbf{\Gamma}\mathbf{r}(t)\right)^{2}+\mu\mathbf{r}^{2}(t)\label{eq:E}
\end{equation}
to be minimized at each time $t$. The first term in $E(t)$ induces
the approximation Equation \eqref{eq:xGammar}, the second term fosters
a low spike rate with spiking distributed over all neurons. The error
function respects causality as it depends implicitly via $\mathbf{x}(t)$
and $\mathbf{r}(t)$ on the past and restrains the dynamics at the
current time $t$ only. Minimizing $E(t)$ at $t$ means that a spike
should be sent by neuron $n$ if $E(t)$ decreases due to this spike.
Comparing $E_{n}(t)$ (spike sending at time $t$ by neuron $n$)
with $E_{0}(t)$ (no spike sending) yields
\begin{align}
E_{n}(t) & <E_{0}(t)\label{eq:-1}\\
\left(\mathbf{x}(t)-\mathbf{\Gamma}\mathbf{r}(t)-\mathbf{\Gamma}\hat{\mathbf{e}}_{n}\right)^{2}+\mu\left(\mathbf{r}(t)+\hat{\mathbf{e}}_{n}\right)^{2} & <\left(\mathbf{x}(t)-\mathbf{\Gamma}\mathbf{r}(t)\right)^{2}+\mu\mathbf{r}^{2}(t)\label{eq:}\\
\mathbf{\mathbf{\Gamma}}_{n}\cdot\left(\mathbf{x}(t)-\mathbf{\Gamma}\mathbf{r}(t)\right)-\mu r_{n}(t) & >\frac{\mathbf{\Gamma}_{n}^{2}+\mu}{2},\label{eq:VandTRelation}
\end{align}
where \foreignlanguage{english}{$\hat{\mathbf{e}}_{n}$ denotes the
n-th unit vector, $\hat{\mathbf{e}}_{n}=(0,...,1,0,...)^{T}$ (with
a $1$ in the n-th row), and $\mathbf{\mathbf{\Gamma}}_{n}$ is the
$n$-th column (vector) of the matrix $\mathbf{\Gamma}$, $\mathbf{\mathbf{\Gamma}}_{n}=\mathbf{\Gamma}\hat{\mathbf{e}}_{n}$.}
To obtain the familiar condition $V_{n}(t)>\theta_{n}$ for neuron
$n$ to spike, the variable left hand side of Equation \eqref{eq:VandTRelation}
may be interpreted as membrane potential,
\begin{align}
V_{n}(t) & =\mathbf{\mathbf{\Gamma}}_{n}\cdot\left(\mathbf{x}(t)-\mathbf{\Gamma}\mathbf{r}(t)\right)-\mu r_{n}(t),\label{eq:VnGammaxGammar}\\
\mathbf{V}(t) & =\mathbf{\Gamma}^{T}\left(\mathbf{x}(t)-\mathbf{\Gamma}\mathbf{r}(t)\right)-\mu\mathbf{r}(t),\label{eq:VGammaxGammar}
\end{align}
the right hand side as threshold
\begin{equation}
\theta_{n}=\frac{\mathbf{\Gamma}_{n}^{2}+\mu}{2}.\label{eq:ThresholdDistributedCase}
\end{equation}
We note that we can multiply both sides of the Equation by a factor
and add constant terms, these change the scale of the potential, and
shift the resting membrane potential, the reset and the threshold.
Equation \eqref{eq:VGammaxGammar} yields Equation \eqref{eq:xLvGx}
with the pseudo-inverse of $\mathbf{\Gamma}^{T}$, $\mathbf{L}=(\mathbf{\Gamma}\mathbf{\Gamma}^{T})^{-1}\mathbf{\Gamma}$,
and $\mathbf{\tilde{\Gamma}}=\mathbf{\Gamma}+\mu\mathbf{L}$.

We can now derive the sub-threshold dynamical Equations for $\mathbf{V}(t)$
from those for $\mathbf{x}(t)$ and $\mathbf{r}(t)$:
\begin{align}
\dot{\mathbf{V}}(t) & =\mathbf{\Gamma}^{T}\left(\dot{\mathbf{x}}(t)-\mathbf{\Gamma}\dot{\mathbf{r}}(t)\right)-\mu\dot{\mathbf{r}}(t)\label{eq:-2}\\
 & =\mathbf{\Gamma}^{T}\mathbf{f}(\mathbf{x}(t))-\left(\mathbf{\Gamma}^{T}\mathbf{\Gamma}+\mu\mathbf{1}\right)\left(\mathbf{s}(t)-\lambda_{s}\mathbf{r}(t)\right)+\mathbf{\Gamma}^{T}\mathbf{c}(t),\label{eq:VdotDistributed1}
\end{align}
where $\mathbf{1}$ denotes the $N\times N$ identity matrix. Assuming
that the minimization of Equation \eqref{eq:E} yields small $E(t)$,
we may eliminate the dependence on $\mathbf{x}(t)$ using Equation
\eqref{eq:xGammar},
\begin{equation}
\dot{\mathbf{V}}(t)\approx\mathbf{\mathbf{\Gamma}}^{T}\mathbf{f}(\mathbf{\Gamma}\mathbf{r}(t))-\left(\mathbf{\Gamma}^{T}\mathbf{\Gamma}+\mu\mathbf{1}\right)\left(\mathbf{s}(t)-\lambda_{s}\mathbf{r}(t)\right)+\mathbf{\mathbf{\Gamma}}^{T}\mathbf{c}(t).\label{eq:VdotDistributed2}
\end{equation}
Finally, biological realism and increased stability of numerical simulations
indicate that an additional leak term $-\lambda_{V}\mathbf{V}$ should
be introduced
\begin{equation}
\dot{\mathbf{V}}(t)=-\lambda_{V}\mathbf{V}(t)+\mathbf{\Gamma}^{T}\mathbf{f}(\mathbf{\Gamma}\mathbf{r}(t))-\left(\mathbf{\Gamma}^{T}\mathbf{\Gamma}+\mu\mathbf{1}\right)\left(\mathbf{s}(t)-\lambda_{s}\mathbf{r}(t)\right)+\mathbf{\Gamma}^{T}\mathbf{c}(t).\label{eq:VdotDistributedFinal}
\end{equation}
We now choose the $f_{j}$ as
\begin{align}
f_{j}(x_{1},...,x_{J}) & =-\lambda_{x}x_{j}+\sum_{i=1}^{J}A_{ji}\tanh\left(x_{i}\right),\label{eq:fxDistributed}
\end{align}
such that
\begin{equation}
\dot{\mathbf{V}}(t)=-\lambda_{V}\mathbf{V}(t)+\mathbf{\Gamma}^{T}\mathbf{A}\tanh(\mathbf{\Gamma}\mathbf{r}(t))-\left(\mathbf{\Gamma}^{T}\mathbf{\Gamma}+\mu\mathbf{1}\right)\mathbf{s}(t)+\left(a\mathbf{\Gamma}^{T}\mathbf{\Gamma}+\mu\lambda_{s}\mathbf{1}\right)\mathbf{r}(t)+\mathbf{\Gamma}^{T}\mathbf{c}(t),\label{eq:VdotDistributedFinalFinal}
\end{equation}
where $a=\lambda_{s}-\lambda_{x}$. This yields a spiking neural network
that is suitable for universal computation: Its dynamics can be decoded
via Equation \eqref{eq:xLvGx} (or Equation \eqref{eq:xGammar}) to
resemble those of a $J$-dimensional dynamical system of the form
\begin{equation}
\dot{\mathbf{x}}(t)=-\lambda_{x}\mathbf{x}(t)+\mathbf{A}\tanh\left(\mathbf{x}(t)\right)+\mathbf{c}(t).\label{eq:xdotDistributedFinalFinal}
\end{equation}
Systems of the form Equation \eqref{eq:xdotDistributedFinalFinal}
are known to be suitable for universal computation \citep{JH04,SA09,lukosevicius2012reservoir},
in particular for appropriate $\mathbf{A}$ they can maintain longer-term
fading memory. Since the $\mathbf{x}$ are dynamical quantities linearily
derived from the underlying spiking network, already the underlying
spiking network is suitable for computations.

Furthermore, the structure of Equation \eqref{eq:VdotDistributedFinalFinal}
allows for a straightforward interpretation in biological terms: \foreignlanguage{english}{The
response of neuron $n$'s soma to slow input to its $J$ dendrites
is modeled by the term $\mathbf{\mathbf{\Gamma}}_{n}\cdot\left(\mathbf{A}\tanh(\mathbf{\Gamma}\mathbf{r}(t))\right)$}.
The inputs have non-negligible synaptic time constant (cf.~$\mathbf{r}(t)$),
they are linearly summed and thereafter subjected to a dendritic sublinearity
($\tanh$). The coupling strength of a synaptic connection from neuron
$m$ to the $j$th dendrite of neuron $n$ is given by $\Gamma_{jm}$,
the coupling strength from the $j$th dendrite of neuron $n$ to its
soma is $(\mathbf{\Gamma}^{T}\mathbf{A})_{nj}$. Further fast and
slow inputs arriving near the soma (and thus not subject to a dendritic
non-linearity) are incorporated by the terms $-\left(\mathbf{\Gamma}^{T}\mathbf{\Gamma}+\mu\mathbf{1}\right)\mathbf{s}(t)$
and $\left(a\mathbf{\Gamma}^{T}\mathbf{\Gamma}+\mu\lambda_{s}\mathbf{1}\right)\mathbf{r}(t)$.
Their impact is characterized by the product $\mathbf{\Gamma}^{T}\mathbf{\Gamma}$
of the decoding matrix with itself and the comparably small weight
$\mu$ of the spike frequency penalty term, the positive diagonal
terms incorporate the reset of the neurons after a spike and a slower
recovery.

\subsection{A sufficient condition for the echo state property of the dynamics
Equation (6)}

When does

\begin{equation}
\dot{{\bf x}}(t)=-\lambda_{V}\left[\mathbf{x}(t)\right]_{-}-\lambda_{x}\left[\mathbf{x}(t)\right]_{+}+\mathbf{A}\tanh(\left[\mathbf{x}(t)\right]_{+})+I(t)\label{eq:conp0}
\end{equation}

(Equation (6) of the main text) possess the echo state property? Dynamics
have this property if after sufficiently long time any initial conditions
are washed out and the state of the system is completely determined
by the input. This is definitely the case if the distance between
trajectories decays at least exponentially with a rate independent
of the input \citep{jaeger2001echo}. We will prove the latter for
our dynamics Equation \eqref{eq:conp0}. For this, we will consider
the difference $\mathbf{\Delta}(t)=\mathbf{x}_{1}(t)-\mathbf{x}_{2}(t)$
and the Euclidean distance $\left\Vert \mathbf{\Delta}(t)\right\Vert $
of two trajectories that satisfy Equation \eqref{eq:conp0} and have
different initial conditions $\mathbf{x}_{1}(0)$,\foreignlanguage{english}{\textrm{$\mathbf{x}_{2}(0)$}}
but the same input $I(t)$. We will show that under the condition
$\left\Vert \mathbf{A}\right\Vert <\min(\lambda_{V},\lambda_{x})$,
with $\left\Vert \mathbf{A}\right\Vert $ being the spectral norm
(the largest singular value) of $\mathbf{A}$, an inequality $\dot{\left\Vert \mathbf{\Delta}(t)\right\Vert }\leq-\epsilon\left\Vert \mathbf{\Delta}(t)\right\Vert $
holds for some $\epsilon>0$ (as usual the dot denotes the temporal
derivative of the entire expression below, here $\left\Vert \mathbf{\Delta}(t)\right\Vert $).

We start with the expression $\frac{1}{2}\dot{\left\Vert \mathbf{\Delta}(t)\right\Vert ^{2}}=\mathbf{\Delta}(t)\dot{\mathbf{\Delta}}(t)$
and replace the right hand side using $\mathbf{\Delta}(t)=\mathbf{x}_{1}(t)-\mathbf{x}_{2}(t)$
and Equation \eqref{eq:conp0}, which leads to
\begin{eqnarray}
\dot{\frac{1}{2}\left\Vert \mathbf{\Delta}(t)\right\Vert ^{2}} & = & \left(\mathbf{x}_{1}-\mathbf{x}_{2}\right)\left(-\lambda_{V}\left(\left[\mathbf{x}_{1}(t)\right]_{-}-\left[\mathbf{x}_{2}(t)\right]_{-}\right)-\lambda_{x}\left(\left[\mathbf{x}_{1}(t)\right]_{+}-\left[\mathbf{x}_{2}(t)\right]_{+}\right)\right)\nonumber \\
 &  & +\left(\mathbf{x}_{1}-\mathbf{x}_{2}\right)\left(\mathbf{A}\left(\tanh(\left[\mathbf{x}_{1}(t)\right]_{+})-\tanh(\left[\mathbf{x}_{2}(t)\right]_{+})\right)+I(t)-I(t)\right).\label{eq:conp1}
\end{eqnarray}

To proceed we use the three inequalities $\mathbf{xy}\leq\left\Vert \mathbf{x}\right\Vert \left\Vert \mathbf{y}\right\Vert $,
$\left\Vert \mathbf{A}\mathbf{x}\right\Vert \leq\left\Vert \mathbf{A}\right\Vert \left\Vert \mathbf{x}\right\Vert $
and $\left\Vert \tanh(\mathbf{x})-\tanh(\mathbf{y})\right\Vert \leq\left\Vert \mathbf{x-y}\right\Vert $,
which allow to estimate
\begin{equation}
\left(\mathbf{x}_{1}-\mathbf{x}_{2}\right)\mathbf{A}\left(\tanh(\left[\mathbf{x}_{1}(t)\right]_{+})-\tanh(\left[\mathbf{x}_{2}(t)\right]_{+})\right)\leq\left\Vert \mathbf{x}_{1}-\mathbf{x}_{2}\right\Vert \left\Vert \mathbf{A}\right\Vert \left\Vert \left[\mathbf{x}_{1}(t)\right]_{+}-\left[\mathbf{x}_{2}(t)\right]_{+}\right\Vert .\label{eq:conp2}
\end{equation}

We now simplify the right hand side of the inequality further. For
this we use that for every pair of real valued vectors $\mathbf{x}$
and $\mathbf{y}$ we have $\left[\mathbf{x}\right]_{\pm}\left[\mathbf{y}\right]_{\pm}\geq0$,
$\left[\mathbf{x}\right]_{\pm}\left[\mathbf{y}\right]_{\mp}\leq0$
and $\left[\mathbf{x}\right]_{\pm}\left[\mathbf{x}\right]_{\mp}=0$,
since every element of $\left[\mathbf{x}\right]_{+}$ is larger/equal
zero while every element of $\left[\mathbf{x}\right]_{-}$ is smaller/equal
zero, and elements which are nonzero in $\left[\mathbf{x}\right]_{+}$
are zero in $\left[\mathbf{x}\right]_{-}$ and vice versa. With this
we get
\begin{eqnarray}
\left\Vert \mathbf{x}_{1}-\mathbf{x}_{2}\right\Vert ^{2} & = & \left\Vert \left[\mathbf{x}_{1}(t)\right]_{+}-\left[\mathbf{x}_{2}(t)\right]_{+}\right\Vert ^{2}+\left\Vert \left[\mathbf{x}_{1}(t)\right]_{-}-\left[\mathbf{x}_{2}(t)\right]_{-}\right\Vert ^{2}\nonumber \\
 &  & -2\left[\mathbf{x}_{1}(t)\right]_{+}\left[\mathbf{x}_{2}(t)\right]_{-}-2\left[\mathbf{x}_{2}(t)\right]_{+}\left[\mathbf{x}_{1}(t)\right]_{-}\nonumber \\
 & \geq & \left\Vert \left[\mathbf{x}_{1}(t)\right]_{+}-\left[\mathbf{x}_{2}(t)\right]_{+}\right\Vert ^{2},\label{eq:conp3}
\end{eqnarray}

\textcolor{black}{since $-\left[\mathbf{x}_{1}(t)\right]_{+}\left[\mathbf{x}_{2}(t)\right]_{-}-\left[\mathbf{x}_{2}(t)\right]_{+}\left[\mathbf{x}_{1}(t)\right]_{-}\geq0$.
The result can be used to bound the right hand side of Equation \eqref{eq:conp2}
by a simpler expression,
\begin{equation}
\left(\mathbf{x}_{1}-\mathbf{x}_{2}\right)\mathbf{A}\left(\tanh(\left[\mathbf{x}_{1}(t)\right]_{+})-\tanh(\left[\mathbf{x}_{2}(t)\right]_{+})\right)\leq\left\Vert \mathbf{x}_{1}-\mathbf{x}_{2}\right\Vert ^{2}\left\Vert \mathbf{A}\right\Vert .\label{eq:conp2-1}
\end{equation}
}

\textcolor{black}{Now we assume $\left\Vert \mathbf{A}\right\Vert <\min(\lambda_{V},\lambda_{x})$
such that we can write $\lambda_{V}=\epsilon_{V}+\left\Vert \mathbf{A}\right\Vert $
and $\lambda_{x}=\epsilon_{x}+\left\Vert \mathbf{A}\right\Vert $
with $\epsilon_{V}>0$ and $\epsilon_{x}>0$. Using this in Equation
\eqref{eq:conp1} yields}

\begin{eqnarray}
\dot{\frac{1}{2}\left\Vert \mathbf{\Delta}(t)\right\Vert ^{2}} & = & -\left(\epsilon_{V}+\left\Vert \mathbf{A}\right\Vert \right)\left(\mathbf{x}_{1}-\mathbf{x}_{2}\right)\left(\left[\mathbf{x}_{1}(t)\right]_{-}-\left[\mathbf{x}_{2}(t)\right]_{-}\right)\nonumber \\
 &  & -\left(\epsilon_{x}+\left\Vert \mathbf{A}\right\Vert \right)\left(\mathbf{x}_{1}-\mathbf{x}_{2}\right)\left(\left[\mathbf{x}_{1}(t)\right]_{+}-\left[\mathbf{x}_{2}(t)\right]_{+}\right)\nonumber \\
 &  & +\left(\mathbf{x}_{1}-\mathbf{x}_{2}\right)\mathbf{A}\left(\tanh(\left[\mathbf{x}_{1}(t)\right]_{+})-\tanh(\left[\mathbf{x}_{2}(t)\right]_{+})\right)\nonumber \\
 & = & -\epsilon_{V}\left(\mathbf{x}_{1}-\mathbf{x}_{2}\right)\left(\left[\mathbf{x}_{1}(t)\right]_{-}-\left[\mathbf{x}_{2}(t)\right]_{-}\right)-\epsilon_{x}\left(\mathbf{x}_{1}-\mathbf{x}_{2}\right)\left(\left[\mathbf{x}_{1}(t)\right]_{+}-\left[\mathbf{x}_{2}(t)\right]_{+}\right)\nonumber \\
 &  & -\left\Vert \mathbf{A}\right\Vert \left\Vert \mathbf{x}_{1}-\mathbf{x}_{2}\right\Vert ^{2}+\left(\mathbf{x}_{1}-\mathbf{x}_{2}\right)\mathbf{A}\left(\tanh(\left[\mathbf{x}_{1}(t)\right]_{+})-\tanh(\left[\mathbf{x}_{2}(t)\right]_{+})\right),\label{eq:conp3-1}
\end{eqnarray}

and together with Equation \eqref{eq:conp2-1} 
\begin{eqnarray}
\dot{\frac{1}{2}\left\Vert \mathbf{\Delta}(t)\right\Vert ^{2}} & \leq & -\epsilon_{V}\left(\mathbf{x}_{1}-\mathbf{x}_{2}\right)\left(\left[\mathbf{x}_{1}(t)\right]_{-}-\left[\mathbf{x}_{2}(t)\right]_{-}\right)-\epsilon_{x}\left(\mathbf{x}_{1}-\mathbf{x}_{2}\right)\left(\left[\mathbf{x}_{1}(t)\right]_{+}-\left[\mathbf{x}_{2}(t)\right]_{+}\right)\nonumber \\
 &  & -\left\Vert \mathbf{A}\right\Vert \left\Vert \mathbf{x}_{1}-\mathbf{x}_{2}\right\Vert ^{2}+\left\Vert \mathbf{x}_{1}-\mathbf{x}_{2}\right\Vert ^{2}\left\Vert \mathbf{A}\right\Vert \nonumber \\
 & = & -\epsilon_{V}\left(\mathbf{x}_{1}-\mathbf{x}_{2}\right)\left(\left[\mathbf{x}_{1}(t)\right]_{-}-\left[\mathbf{x}_{2}(t)\right]_{-}\right)-\epsilon_{x}\left(\mathbf{x}_{1}-\mathbf{x}_{2}\right)\left(\left[\mathbf{x}_{1}(t)\right]_{+}-\left[\mathbf{x}_{2}(t)\right]_{+}\right).\label{eq:conp3-2}
\end{eqnarray}

Both terms on the right hand side are smaller or equal to zero,
\begin{equation}
\left(\mathbf{x}_{1}-\mathbf{x}_{2}\right)\left(\left[\mathbf{x}_{1}(t)\right]_{\pm}-\left[\mathbf{x}_{2}(t)\right]_{\pm}\right)=\left(\left[\mathbf{x}_{1}(t)\right]_{\pm}-\left[\mathbf{x}_{2}(t)\right]_{\pm}\right)^{2}-\left[\mathbf{x}_{1}(t)\right]_{\pm}\left[\mathbf{x}_{2}(t)\right]_{\mp}-\left[\mathbf{x}_{2}(t)\right]_{\pm}\left[\mathbf{x}_{1}(t)\right]_{\mp}\geq0.\label{eq:conp4}
\end{equation}

We can therefore set $\epsilon=\min(\epsilon_{V},\epsilon_{x})>0$
and simplify
\begin{eqnarray}
\dot{\frac{1}{2}\left\Vert \mathbf{\Delta}(t)\right\Vert ^{2}} & \leq & -\epsilon\left(\mathbf{x}_{1}-\mathbf{x}_{2}\right)\left(\left[\mathbf{x}_{1}(t)\right]_{-}-\left[\mathbf{x}_{2}(t)\right]_{-}\right)-\epsilon\left(\mathbf{x}_{1}-\mathbf{x}_{2}\right)\left(\left[\mathbf{x}_{1}(t)\right]_{+}-\left[\mathbf{x}_{2}(t)\right]_{+}\right)\nonumber \\
 & = & -\epsilon\left\Vert \mathbf{\Delta}(t)\right\Vert ^{2},
\end{eqnarray}

which is equivalent to
\begin{equation}
\dot{\left\Vert \mathbf{\Delta}(t)\right\Vert }\leq-\epsilon\left\Vert \mathbf{\Delta}(t)\right\Vert .
\end{equation}

The distance between different trajectories thus decreases at least
exponentially fast with rate $\epsilon$, for any input. We may conclude
that $\left\Vert \mathbf{A}\right\Vert <\min(\lambda_{V},\lambda_{x})$
provides a sufficient condition for the system to possess the echo-state
property.

\subsection{Comparison of PCSNs with Poisson coding learning networks}

\begin{figure}
\includegraphics[width=1\textwidth]{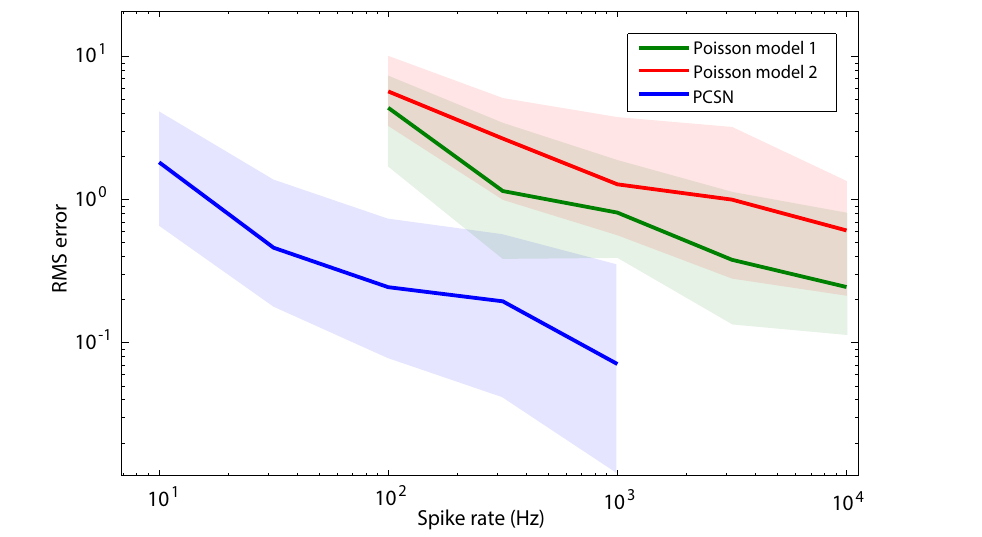}

\protect\caption{\textbf{Comparison of PCSNs and Poisson coding learning networks.}
Error of PCSNs and Poisson coding learning networks with different
spike rates after learning continuous dynamics. The panel shows the
median error to the saw tooth target pattern (cf.~Fig.~3d) during
testing, in equidistant bins of the network spike rate (shaded: intervals
between first and third quartile). The PCSN with its deterministic
spike code reaches the same error level as the networks with simple
Poisson coding with almost two orders of magnitude fewer spikes. \label{fig:SF1}}
\end{figure}

In the following, we compare the performance of PCSNs and Poisson
coding learning networks. To enable a direct comparison, we use PCSNs
with saturating synapses and Poisson coding networks of the same size
and with the same learning rule for the recurrent synapses such that
in the high-rate limit, both network types become equivalent to the
same continuous networ\textcolor{black}{ks. As a specific task for
the comparison, we choose learning of a saw tooth-like signal as displayed
in Fig.~3d. We find that both networks perform better for higher
rates. However, due to their deterministic, precise spike code, the
PCSNs achieve the same error levels with almost two orders of magnitude
smaller rates, cf.~Fig.~A. This is generally a consequence of the
fact that the population coding error in precisely spiking networks
is much smaller than in Poisson coding networks, ref.~\citep{boerlin2013predictive}
shows it to be proportional to $\nicefrac{1}{N}$ (where $N$ is the
number of neurons in the network), while a simple Poisson population
code has precision $\nicefrac{1}{\sqrt{N}}$. However, in our PCSNs
we have additional learning whose consequences on the precision of
the output signal are not easy to determine.}

The Poisson models are setup as follows: We start with the continuous
target dynamics Equation (6) and for simplicity consider $\lambda_{x}=\lambda_{V}$
and $\gamma=\theta$, i.e.
\begin{equation}
\dot{x}_{n}(t)=-\lambda_{V}x_{n}(t)+\sum_{m=1}^{N}A_{nm}\tanh(\left[x_{m}(t)\right]_{+})+I_{\text{e},n}(t).\label{eq:-7}
\end{equation}
The state of the corresponding Poisson unit $n$ shall be characterized
by $u_{n}(t)$; we aim at $u_{n}(t)\approx x_{n}(t)$ for high spike
rates. For the spike generation, we orient at standard models (e.g.~\citep{GK06})
and at keeping the dynamical Equations simple.

As Poisson model 1, we use networks of units with threshold and nonlinear
saturation, specifically unit $n$ has the rate
\begin{equation}
\nu_{n}(t)=s_{0}\tanh\left(\left[u_{n}(t)\right]_{+}\right).\label{eq:-8}
\end{equation}
The constant $s_{0}$ allows to modulate the rate without changing
the dynamics of ${\bf u}$. Given $\nu_{n}(t)$, the unit generates
an inhomogeneous Poisson spike train $s_{n}(t)$ (cf.~Equation (1))
with this rate. The spike train in turn generates postsynaptic inputs
with decay time constant $\lambda_{s}$, as given in Equation (2).
When rescaled with $\lambda_{s}/s_{0}$, the postsynaptic inputs satisfy
for large $s_{0}$ 
\begin{equation}
\frac{\lambda_{s}}{s_{0}}r_{n}(t)\approx\frac{\nu_{n}(t)}{s_{0}}=\tanh\left(\left[u_{n}(t)\right]_{+}\right).\label{eq:-10}
\end{equation}
A network with dynamics
\begin{equation}
\dot{u}_{n}(t)=-\lambda_{V}u_{n}(t)+\sum_{m=1}^{N}A_{nm}\frac{\lambda_{s}}{s_{0}}r_{m}(t)+I_{\text{e},n}(t)\label{eq:-9}
\end{equation}
then approximates the continuous dynamics Equation \eqref{eq:-7}.

As Poisson model 2, we use networks of linear threshold units,
\[
\nu_{n}(t)=s_{0}\left[u_{n}(t)\right]_{+}.
\]
They yield for large $s_{0}$ 
\begin{equation}
\frac{\lambda_{s}}{s_{0}}r_{n}(t)\approx\left[u_{n}(t)\right]_{+}\label{eq:-10-1}
\end{equation}
and
\begin{equation}
\dot{u}_{n}(t)=-\lambda_{V}u_{n}(t)+\sum_{m=1}^{N}A_{nm}\tanh\left(\frac{\lambda_{s}}{s_{0}}r_{m}(t)\right)+I_{\text{e},n}(t)\label{eq:-9-1}
\end{equation}
for the network dynamics approximating Equation \eqref{eq:-7}. We
note that this model also satisfies a decoding Equation analogous
to Equation (9), $\frac{\lambda_{s}}{s_{0}}r_{n}(t)\approx\left[x_{n}(t)\right]_{+}$.

We change the threshold $\theta$ (PCSNs) and the base rates $s_{0}$
(Poisson networks) to generate networks with different rates. For
PCSNs the remaining parameters are adapted such that the corresponding
continuous network is also given by Equation \eqref{eq:-7}. For each
parameter value we train $75$ networks with different random topology
and different initial conditions. We thereafter compute the actually
generated average spike rates within log-scale equidistant bins. Further,
we compute the root mean squared (RMS) error between the desired signal
and the signal generated during testing. Fig.~A displays the median
and the first and third quartiles of the error versus the average
rate in double logarithmic scale.

\subsection{Robustness against noise}

\begin{figure}

\subsection*{\protect\includegraphics[width=1\textwidth]{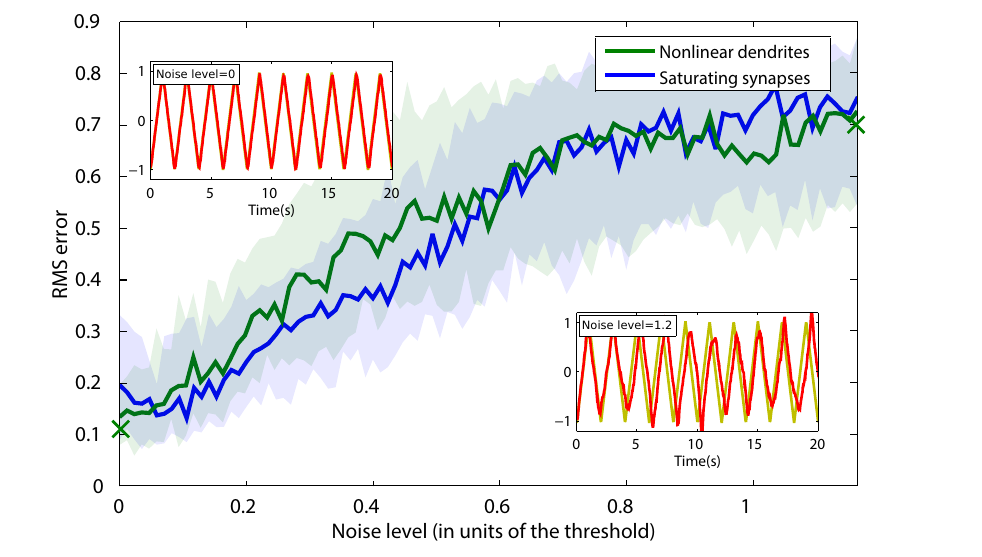}}

\protect\caption{\textbf{Robustness of \mbox{PCSNs} against noise.} The figure shows
the median RMS error between the output of PCSNs and the saw tooth
target pattern during testing, versus the noise level (shaded: intervals
between first and third quartile). The noise level is given in terms
of the standard deviation generated by a purely noise-driven subthreshold
membrane potential (Ornstein-Uhlenbeck process) with membrane time
constant $\lambda_{V}$, in multiples of the threshold. The insets
display the testing phase for two examples using the setup with nonlinear
dendrites (crosses in the main plot denote the corresponding noise
and error levels).\label{fig:SF2}}
\end{figure}

PCSN learning is robust against noise. Fig.~B shows this by example
of the learning of the saw tooth pattern (cf.~Figs.~3d, A), both
for PCSNs with saturating synapses and nonlinear dendrites. In each
time step of the Euler-Maruyama integration, Gaussian noise is added.
The noise level is given in terms of the standard deviation generated
by a purely noise-driven subthreshold membrane potential (Ornstein-Uhlenbeck
process) with membrane time constant $\lambda_{V}$, in multiples
of the threshold (which equals $\theta/2$ in the case of saturating
synapses and $\theta$ in the case of non-linear dendrites), i.e.~$\text{noise-level}:=\frac{\sigma_{\eta}}{\sqrt{2\lambda_{V}}}\frac{\sqrt{(1-e^{-2})}}{\text{threshold}}$.
The error is determined as RMS error between the desired signal and
the signal generated during testing.

\subsection{Robustness of PCSNs with nonlinear dendrites against structural perturbations}

\begin{figure}

\subsection*{\protect\includegraphics[width=1\textwidth]{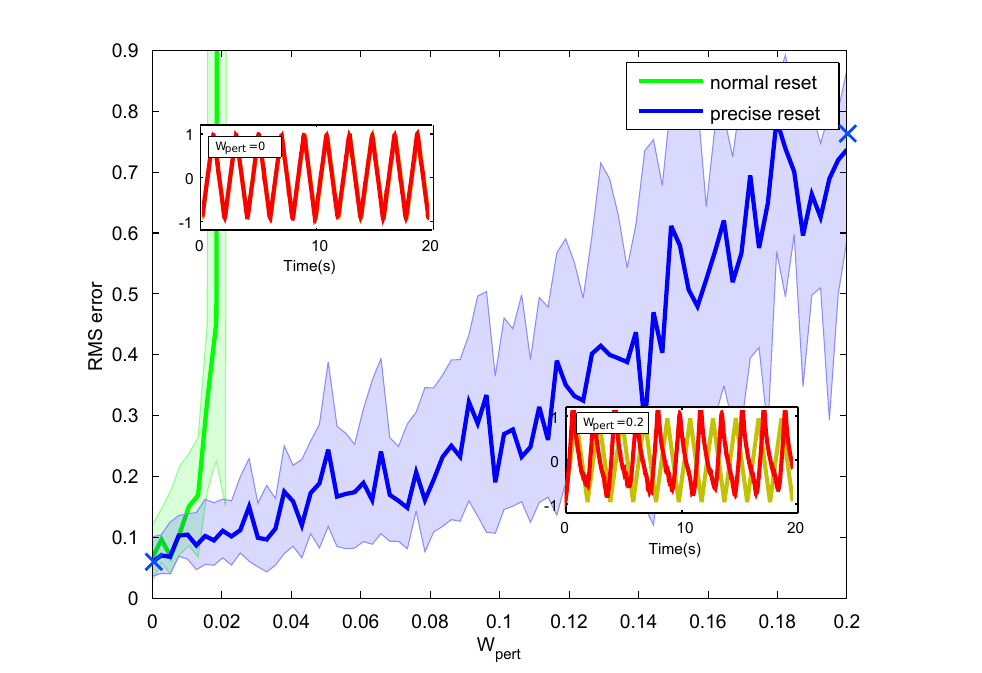}}

\protect\caption{\textbf{Robustness of \mbox{PCSNs} against structural perturbations
of dendritic coupling.} The figure shows performance of native PCSNs
(networks without precise reset, green) and for PCSNs where neurons
are reset to a fixed membrane potential after spike generation (networks
with precise reset, blue). Displayed is the median RMS error between
the output of PCSNs and the saw tooth target pattern during testing
versus the standard deviation $W_{\text{pert}}$ of a multiplicative
Gaussian perturbation in the connectivity (shaded: intervals between
first and third quartile). The insets display the testing phase for
two examples using the setup with precise reset (crosses in the main
plot denote the corresponding $W_{\text{pert}}$ and error levels).
While the native model does not show successful learning for perturbations
larger than $2\%$ of the connection weights, networks with precise
reset generate a recognizable sawtooth pattern as learned output even
for perturbations of $20\%$ (inset at lower right). The error generated
by networks with precise reset increases gradually with perturbation
strength, while there is an abrupt change for the native networks.}
\end{figure}

\begin{figure}

\subsection*{\protect\includegraphics[width=1\textwidth]{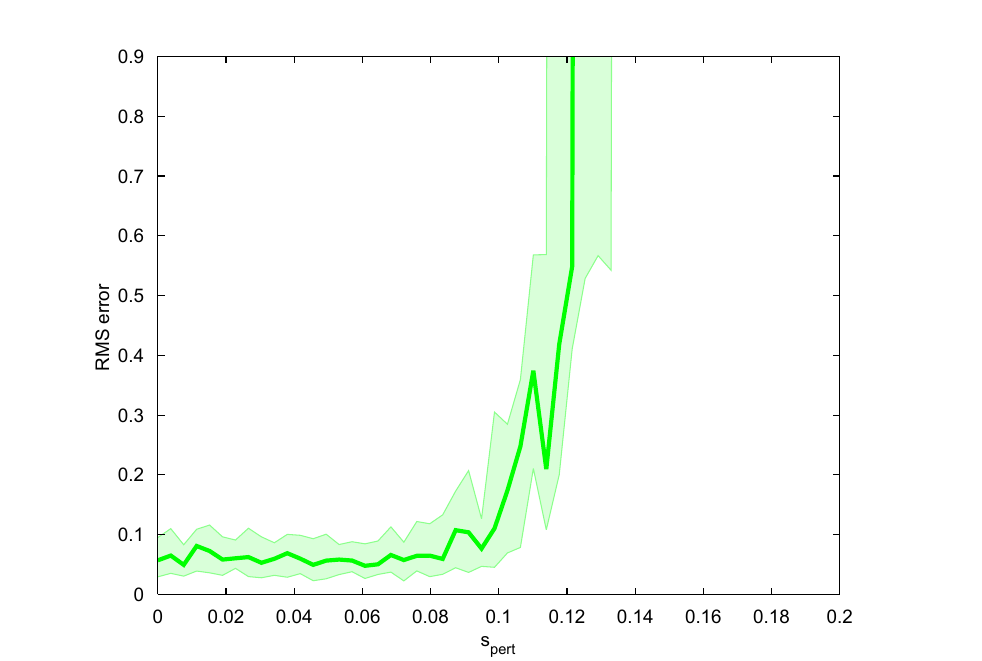}}

\protect\caption{\textbf{Robustness of \mbox{PCSNs} against reduction of fast couplings.}
The panel shows the median RMS error between the output of PCSNs and
the saw tooth target pattern during testing versus the size $s_{\text{pert}}$
of the multiplicative reduction of the fast connections (shaded:
intervals between first and third quartile). For $s_{\text{pert}}<10\%$
the error increases only slightly with increasing $s_{\text{pert}}$.
A further increase of $s_{\text{pert}}$ leads to a strong, rapid
increase in the error and the PCSN is soon not able to learn the pattern
anymore.\label{fig:SF2-1-1}}
\end{figure}

In CSNs with nonlinear dendrites, the optimal strengths of the couplings
from other neurons to the nonlinear dendrites and the fast couplings
are independent of the parameters of the encoded nonlinear dynamical
system Equation \eqref{eq:xdotDistributedFinalFinal}. Since deviations
from the optimal values in these couplings in general do not imply
a simple change in the encoded dynamics but impair the coding scheme,
we here investigate robustness of PCSN-learning against them.

The optimal coupling strength from neuron $m$ to the $j$th nonlinear
dendrite of neuron $n$ is $W_{njm}=\Gamma_{jm}$. Here we test the
robustness of the learning scheme against deviations from the optimal
couplings. We find that PCSN learning is robust, if we modify the
neuron model to have a ``precise reset'', i.e.~the reset is always
to the fixed value $-\theta$ ($-2\theta$ below threshold $\theta$),
even if fast excitation to a suprathreshold potential caused the spike
(Fig.~C). In the native model the reset has fixed size $-2\theta$
such that the membrane potential would be reset to a value larger
than $-\theta$ after a suprathreshold excitation. We note that the
reset to a fixed value may also be biologically more plausible than
a reset of fixed size.

Fig.~C shows this by example of the learning of the saw tooth pattern
(cf.~Figs.~3d, A). The $W_{njm}$ are perturbed proportionally to
the strength of their optimal values, the perturbed couplings are
given by $W_{njm}=\Gamma_{jm}\left(1+W_{\text{pert}}\Xi_{jm}^{n}\right)$\foreignlanguage{english}{\textrm{,
}}where the \foreignlanguage{english}{\textrm{$\Xi_{jm}^{n}$}} are
independently drawn from a Gaussian distribution with mean zero and
variance one. We adopt the interpretation of the PCSNs as networks
with plastic recurrent connections: The outputs, the output weight
updates, the inverse correlation matrices and the updates of the dendrite-to-soma
weights $D_{nj}$ are computed using Equations (14), (18), (19) and
(22) with unperturbed readouts $\tilde{r}_{j}(t)=\tanh\left(\sum_{m=1}^{N}\Gamma_{jm}r_{m}(t)\right)$.
We note that updating the $D_{nj}$ is not equivalent to a static
feedback of the updated overall network readout anymore, since the
latter does not contain the perturbed $W_{njm}$.

Our findings raise the question why the networks with precise reset
are much more stable to perturbations in the dendrites. We find that
the instability in the simulations of the conventional model is due
to an explosion of the spike rates, such that every neuron spikes
once at every simulated time step. This is due to the ``ping pong
effect'' already described in ref.~\citep{boerlin2013predictive}:
Neurons that spike due to excitation from fast connections generate
further suprathreshold excitation and in the end all neurons have
spiked within a single step. In \citep{boerlin2013predictive} this
problem is solved using a higher value of $\mu$. We find for our
simulations with perturbed $W_{njm}$ that a fine tuning of $\mu$
is required to prevent a breakdown of the learning. In contrast, the
precise reset solves this problem robustly. This is a consequence
of the fact that on the one hand the precise reset yields a membrane
potential that is further away from the threshold and thus reduces
the chance of re-excitation of a neuron that has recently spiked ($\mu$
has an in principle similar effect). On the other hand, the difference
from the theory arises only after suprathreshold excitation, the change
in the precise spiking dynamics is small and the spike coding scheme
is preserved.

The optimal strength of a fast coupling from neuron $m$ to neuron
$n$ is \foreignlanguage{english}{\textbf{$U_{nm}=\sum_{j=1}^{J}\G_{jn}\G_{jm}+\mu\delta_{nm}$}}
(which is independent of $N$ for fixed neuron threshold). In ``balanced
state'' irregular spiking networks, such recurrent connections may
be expected to be much smaller, since they scale with $1/\sqrt{N}$
for fixed neuron thresholds \citep{RRBHPRH10}. We therefore test
the dependence of the PCSNs on these connections by a multiplicative
weakening by a factor $1-s_{\text{pert}}$, $U_{nm}^{\text{pert}}=(1-s_{\text{pert}})U_{nm}$
for $n\neq m$ (resets are kept $U_{nn}^{\text{pert}}=U_{nn}$). Fig.~D
shows that PCSNs are robust against this: The learning capabilities
are conserved, if $s_{\text{pert}}<10\%$. The figure also indicates
that the fast connections are important for PCSN functionality, since
the learning abilities are quickly lost as $s_{\text{pert}}$ increases
beyond this range.

\subsection{Comparison of network sizes and the spike rate-network size trade-off}

\begin{figure}
\includegraphics[width=1\textwidth]{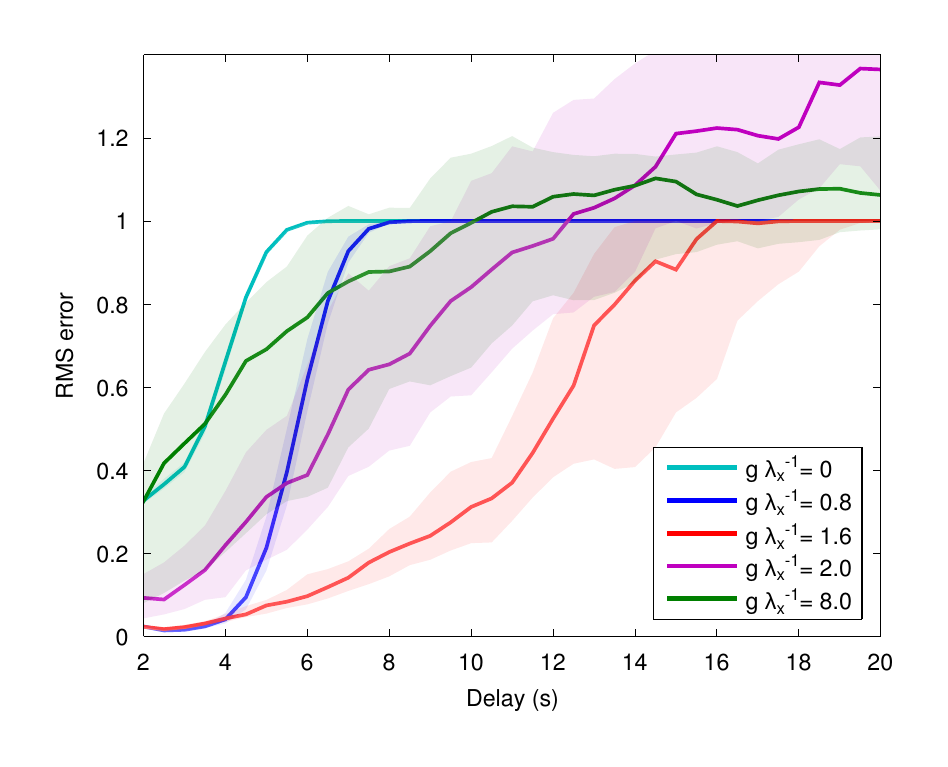}

\protect\caption{\textbf{Memory duration in CSNs with different recurrent coupling
strengths.} Supporting figure to Fig.~4a-c, displaying a direct comparison
between multiple error vs.~reaction delay traces. A disconnected
network, $\text{g}\lambda_{x}^{-1}=0$, has comparably short memory.
Increase of connection strength leads to an increase of memory duration
(cf.~the trace for $\text{g}\lambda_{x}^{-1}=0.8$). Memory is most
persistent around $\text{g}\lambda_{x}^{-1}=1.6$, and decreases for
larger coupling strengths, as expected for systems where the dynamics
become more and more chaotic ($\text{g}\lambda_{x}^{-1}=2$, $\text{g}\lambda_{x}^{-1}=8$).\label{fig:SF3}}
\end{figure}

\begin{figure}
\includegraphics[width=1\textwidth]{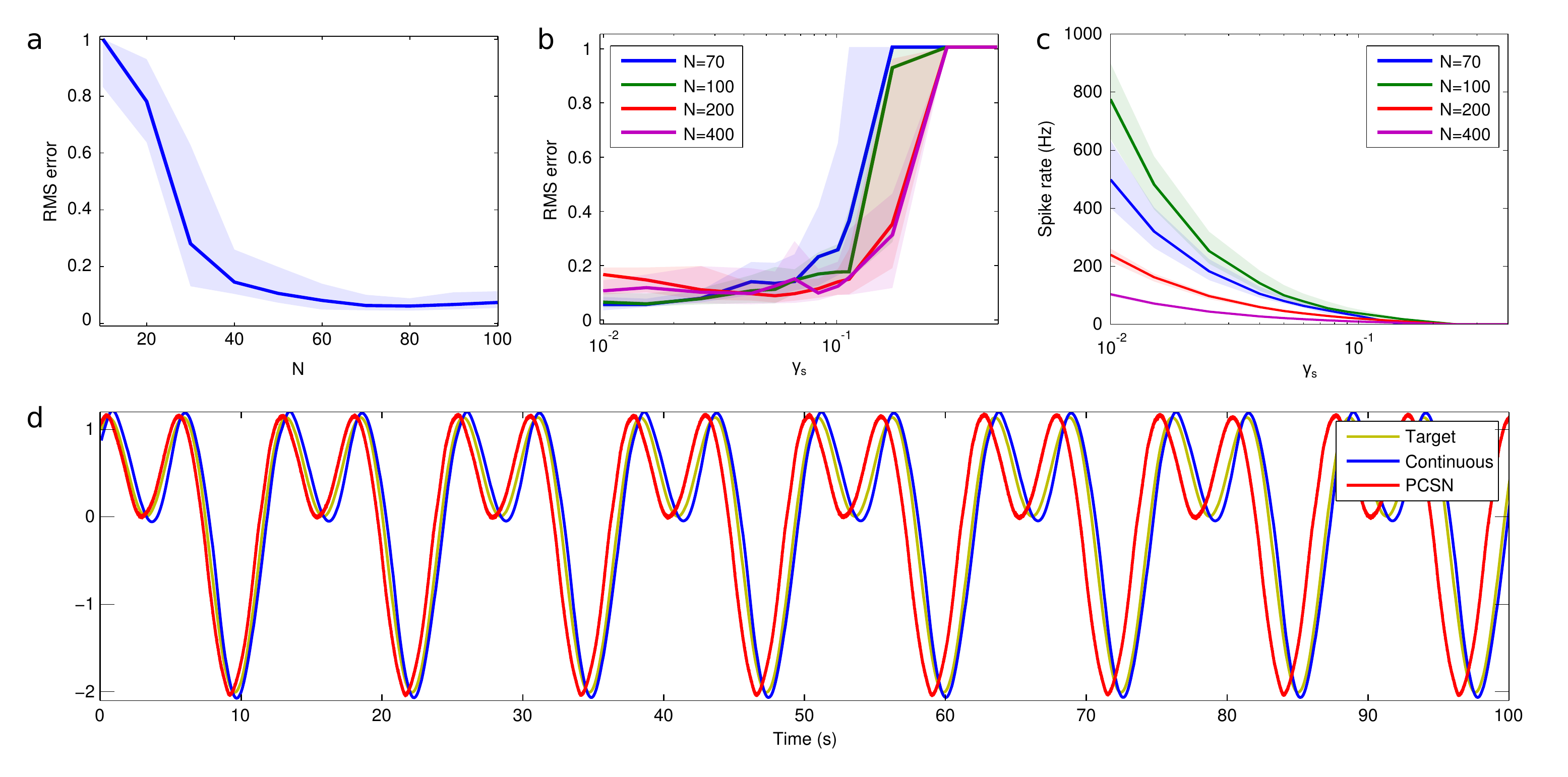}

\protect\caption{\textbf{Network size and PCSN spiking rate trade-off. }(a): RMS errors
of continuous rate networks of different size, after training the
camel's hump task (cf.~Fig.~3e). The networks generate the pattern
well for network sizes larger approximately $N=40$. (b,c): RMS errors
of PCSNs with nonlinear dendrites, after training the camel's hump
task. The panels display the trade-off between network size and spike
frequency: For $N=200$, a small error can be achieved with a single
neuron spike rate of $100$Hz (cf.~RMS error at $\gamma_{s}\approx0.1$
displayed in (b) and rate at $\gamma_{s}\approx0.1$ displayed in
(c)). Networks with $N=400$ need only a spike rate of $10$Hz. Smaller
networks (e.g.~$N=70$) need higher spike rates but reach the same
error levels. (We note that for small networks we observe a nonmonotonic
dependence of the spike rate on the network size for constant $\gamma_{s}$.)
(d): Example dynamics. The PCSN signal (red trace, $\gamma_{s}=0.01$,
$N=70$, $J=50$) approximates the target function (yellow trace)
similarly well as a continuous rate neuron network of similar size
(blue trace, $N=50$).\label{fig:SF4}}
\end{figure}

We illustrate that for PCSNs, network sizes comparable to those of
continuous rate networks solving the same task with FORCE learning
can be sufficient. Further, we show that there is a trade-off between
network size and spike rate of individual neurons. As example we use
the ``camel's hump'' task (cf. Fig. 3e). Fig.~Fa shows that continuous
networks Equation (6) can learn the signal well for about $N>40$,
we use $N=50$ as a reference. We compare with PCSNs with nonlinear
dendrites that encode a system Equation (6) with the same parameters
(in particular $J=50$), and are trained to solve the same task. We
compare the RMS error and spiking frequencies for different PCSN sizes
$N$ and $\gamma_{s}$, which regulates the threshold of the neurons
(cf.~Methods and Equation \eqref{eq:ThresholdDistributedCase}).
Fig.~Fb,c shows the trade-off between the number of neurons in the
PCSNs and their individual spiking frequency: For $\gamma_{s}\approx0.1-0.15$
only the networks with $N=200$ and $N=400$ learn the task reliably
(panel (b)), the neurons adopt a mean spike rate of about or smaller
$100$Hz. For sufficiently small $\gamma_{s}$ also smaller networks
learn the task well, but all networks generate a higher spike frequency.

\subsection{Error evolution for longer times}

\begin{figure}
\includegraphics[width=1\textwidth]{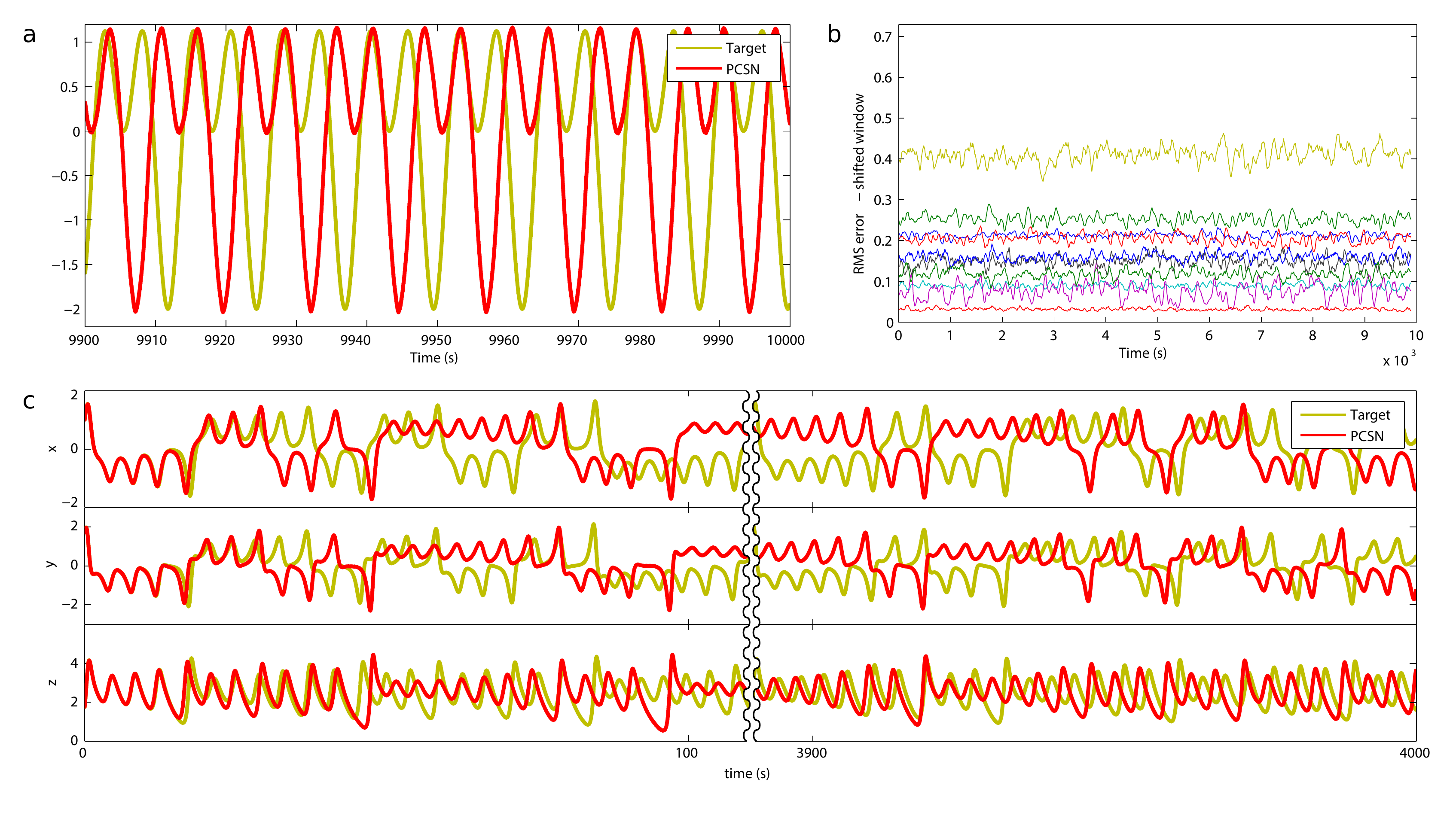}

\protect\caption{\textbf{Long term evolution.} (a): PCSN generated camels's hump signal
and target (Figs.~3e, F) after long times: the signal is phase-shifted
compared to the target but otherwise not noticeably changed. Panel
(b) quantifies this observation by plotting the RMS error, corrected
for a possible phase shift to the target, against time. Displayed
are several learning trials with different random initial connectivity.
The error is stationary, constant except for fluctuations. (c): Lorenz
attractor signal of Fig.~3f-h, dynamics of $x,y,z$ vs.~time. Also
after long times, the PCSN (red trace) in general generates qualitatively
the same dynamics as the target Lorenz system (yellow trace).\label{fig:SF5}}
\end{figure}

We do not observe lasting changes of the error in long term simulations.
For periodic signals, there is an inevitable phase shift. It originates
from the small error between the period of the desired and the learned
signal; this error accumulates over time. Apart from that, tested
features such as the deviation from the desired dynamics (RMS error)
and the spike frequency are remarkably constant. Fig.~Ga,b illustrates
and quantifies this for the camel's hump task (Fig.~3e). Fig.~Ga
displays the continued desired dynamics and the occurring phase shift
for longer recall durations. Fig.~Gb shows that the RMS error is
stationary, approximately constant over time, if one corrects for
the shift. For the Lorenz attractor (Fig.~3f-h), the teacher and
student trajectories quickly depart from each other after the end
of learning. The spiking network nevertheless continues to generate
dynamics that after decoding agree with the dynamics of a Lorenz system.
Fig.~Gc shows this for longer times. As a quantitative check, the
tent map Fig.~3h relating subsequent local maxima in the z-coordinate
shows the good agreement with the tent-map generated by the teacher
dynamics also for long times. In the displayed simulation, the Lorenz
dynamics deviate three times from the desired dynamics, which leads
to the six outliers in Fig. 3h. However, the dynamics return every
time to the desired dynamics such that the errors just generate single
outlier pairs in the tent map and the qualitative dynamics agree with
those of the Lorenz system still after $4000$s.

\section{Supporting methods}

\subsection{Details on the supporting figures}

The parameters of the different simulations are given in Table \ref{tab:ParametersSatSynSUP}
for simulations using saturating synapses and in Table \ref{tab:ParametersNonDenSUP}
for simulations using nonlinear dendrites. Further parameters and
details about the figures and simulations are given in the following
paragraphs.

If not mentioned otherwise, for all simulations we use $g=1.5\frac{1}{\text{s}}$,
$p=0.1$, $\tilde{w}^{f}=1\frac{1}{\s}$, $\tilde{w}^{i}=1\frac{1}{\s}$,
$\Delta t=0.01\text{s}$, $\gamma=\theta$ and $\sigma_{\eta}=0\frac{1}{\sqrt{\s}}$.

\begin{table}
\begin{tabular}{|l||c|c|c|c|c|c|c|c|}
\hline 
Sat. syn. & N & $\alpha$ & dt & $T_{t}$ & $\lambda_{s}^{-1}$ & $\lambda_{V}^{-1}$ & $\text{V}_{r}$ & $\theta$\tabularnewline
\hline 
\hline 
Fig.~A (PCSN) & $50$ & $0.01$ & $0.001$ms & $10.5$s & $100$ms & $1$s & $0.9\theta$ & {\scriptsize{}swept}\tabularnewline
\hline 
Fig.~B & $100$ & $0.1$ & $0.1$ms & $30$s & $100$ms & $50$ms & $0.9\theta$ & $0.05$\tabularnewline
\hline 
\end{tabular}

\protect\caption{Parameters used in the different figures for simulations of networks
with saturating synapses.\label{tab:ParametersSatSynSUP}}
\end{table}

\begin{table}
\begin{tabular}{|l||c|c|c|c|c|c|c|c|c|c|}
\hline 
Nonlin. dendr. & N & J & $\alpha$ & $\gamma_{s}$ & dt & $T_{t}$ & $\mu$ & $\lambda_{s}^{-1}$ & $\lambda_{V}^{-1}$ & $\text{a}$\tabularnewline
\hline 
\hline 
Fig.~B & $200$ & $100$ & $0.1$ & $0.05$ & $0.1$ms & $30$s & $1/N^{2}$ & $100$ms & $50$ms & $\lambda_{s}$$-1$Hz\tabularnewline
\hline 
Fig.~C & $200$ & $100$ & $0.1$ & $0.05$ & $0.1$ms & $30$s & $1/N^{2}$ & $100$ms & $50$ms & $\lambda_{s}$$-1$Hz\tabularnewline
\hline 
Fig.~D & $200$ & $100$ & $0.1$ & $0.05$ & $0.1$ms & $30$s & $1/N^{2}$ & $100$ms & $50$ms & $\lambda_{s}$$-1$Hz\tabularnewline
\hline 
Fig. Fb, c & {\scriptsize{}swept} & $50$ & $1$ & {\scriptsize{}swept} & $0.1$ms & $100.5$s & $0$ & $100$ms & $100$ms & $\lambda_{s}$$-1$Hz\tabularnewline
\hline 
Fig. Fd & $70$ & $50$ & $1$ & {\scriptsize{}0.01} & $0.1$ms & $100.5$s & $0$ & $100$ms & $100$ms & $\lambda_{s}$$-1$Hz\tabularnewline
\hline 
Fig. Ga, b & $70$ & $50$ & $1$ & {\scriptsize{}0.01} & $0.1$ms & $100.5$s & $0$ & $100$ms & $100$ms & $\lambda_{s}$$-1$Hz\tabularnewline
\hline 
\end{tabular}

\protect\caption{Parameters used in the different figures for simulations of networks
with nonlinear dendrites. The parameter $a=\lambda_{s}-\lambda_{x}$
is given in terms of $\lambda_{s}$ and $\lambda_{x}$\label{tab:ParametersNonDenSUP}}
\end{table}

\subsubsection*{Figure A}

The signal has period $2\s$ and amplitude $10$. The parameters of
the Poisson networks are $N=50$, $\alpha=0.01$, $g=1.5\frac{1}{\s}$,
$T_t=10.5$s, $\lambda_{s}^{-1}=100$ms, $\lambda_{V}^{-1}=1$s, $dt=0.1/s_0$,
the sparse matrix $\mathbf{A}$ has a fraction $p=0.1$ of nonzero
entries, which are drawn from a Gaussian distribution with zero mean
and variance $\frac{g^{2}}{pN}$. $s_0$ is swept between $50\frac{1}{s}$
and $21544\frac{1}{s}$ (values in the different sweeps: $s_0=$ $50\frac{1}{\s}$,
$100\frac{1}{\s}$, $150\frac{1}{\s}$, $200\frac{1}{\s}$, $215\frac{1}{\s}$,
$464\frac{1}{\s}$, $1000\frac{1}{\s}$, $2150\frac{1}{\s}$, $4641\frac{1}{\s}$,
$10000\frac{1}{\s}$, $21544\frac{1}{\s}$). The PCSN has saturating
synapses. $\theta$ is swept between $0.5$ and $0.01$ (specific
values of the sweep: $\theta=$$0.5$, $0.4$, $0.3$, $0.2$, $0.1$,
$0.09$, $0.08$, $0.07$, $0.06$, $0.05$, $0.04$, $0.03$, $0.02$,
$0.01$). We compute the RMS error between the signal and the target
in the first $10.5\text{s}$ after training. The error is computed
using a normalized version of the signal, where the amplitude is set
to $1$.

\subsubsection*{Figure B}

The signal has period $2\s$ and amplitude $15$ (normalized to one
in the figure). We take medians and quartiles over $50$ trials. The
noise level is given in terms of the standard deviation generated
by a purely noise-driven subthreshold membrane potential (Ornstein-Uhlenbeck
process) with membrane time constant $\lambda_{V}$, in multiples
of the threshold (which equals $\theta/2$ in the case of saturating
synapses and $\theta$ in the case of non-linear dendrites), i.e.~$\text{noise-level}:=\frac{\sigma_{\eta}}{\sqrt{2\lambda_{V}}}\frac{\sqrt{(1-e^{-2})}}{\text{threshold}}$.
Plotted are the median of the RMS error (shaded: intervals between
first and third quartile) between the signal and the target in the
first 30s after training. The error is computed using the normalized
version of the signal, where the amplitude is set to $1$. The sweep
covers $101$ equidistant values of $\sigma_{n}$ from $0\frac{1}{\sqrt{s}}$
to $0.01\frac{1}{\sqrt{s}}$ in the case of non-linear dendrites and
$101$ equidistant values of $\sigma_{n}$ from $0\frac{1}{\sqrt{s}}$
to $0.2\frac{1}{\sqrt{s}}$ in the case of saturating synapses.

\subsubsection*{Figure C}

The signal has period $2\s$ and amplitude $10$ (normalized to one
in the figure). We take medians and quartiles over $50$ trials. We
use the Euler method to integrate the differential Equations.

Plotted are the median of the RMS error (shaded: intervals between
first and third quartile) between the signal and the target in the
first 30s after training. The error is computed using the normalized
version of the signal, where the amplitude is set to $1$. The sweep
covers $80$ equidistant values of $W_{\text{pert}}$ from $0$ to
$0.2$.

For simulations that showed pathological spiking (more than 200 spikes
per time step in the numerical simulation) we assigned an infinite
error.

\subsubsection*{Figure D}

The signal has period $2\s$ and amplitude $10$ (normalized to one
in the figure). We take medians and quartiles over $50$ trials. We
use the Euler method to integrate the differential Equations.

Plotted are the median of the RMS error (shaded: intervals between
first and third quartile) between the signal and the target in the
first 30s after training. The error is computed using the normalized
version of the signal, where the amplitude is set to $1$. The sweep
covers $80$ equidistant values of $s_{\text{pert}}$ from $0$ to
$0.3$ (displayed is the range up to $0.2$).

For simulations that showed pathological spiking (more than 200 spikes
per time step in the numerical simulation) we assigned an infinite
error.

\subsubsection*{Figure E}

The parameters are as in Fig.~4a-c, lower panels, see ``Figure details''
in the main text.

\subsubsection*{Figure F}

Fig. Fa: The continuous networks obey Equation (11) (Equation \eqref{eq:xdotDistributedFinalFinal}),
they are endowed with the FORCE learning rule. The parameters of the
network are $\lambda_{x}=1\frac{1}{s}$, $dt=0.01\text{s}$, and the
sparse matrix $\mathbf{A}$ has a fraction $p=0.1$ of nonzero entries,
which are drawn from a Gaussian distribution with zero mean and variance
$\frac{g^{2}}{pN}$ with $g=1.5\frac{1}{s}$. The task from Fig.~3e
serves as target signal. The learning rate is $\alpha=1$, the learning
time is $T_{t}=100.5$s. The network size $N$ is swept from $10$
to $100$ in steps of $10$. The figure shows the median of the RMS
error (shaded: intervals between first and third quartile) between
the signal and the target in the first $10$s after training. The
statistics are based on $100$ trials per value of $N$.

Fig. Fb,c: We use PCSNs with nonlinear dendrites, which encode continuous
dynamics as generated by the rate networks in panel (a). $\gamma_{s}$
is swept over $0.4,0.25,0.15,0.1,0.09,0.075,0.06,0.05,0.04,$ $0.025,0.015,0.01$
and networks of size $N$ are plotted for $N=70,100,200,400$. Fig.~Fb
shows the median of the RMS error (shaded: interval between first
and third quartile) between the signal and the target in the first
$10$s after training. Fig.~Fc shows the median of the mean spike
rate per neuron (shaded: intervals between first and third quartile).
The statistics are based on $20$ trials per parameter combination.

Fig. Fd: The plot shows example patterns generated by continuous networks
as used in panel (a) (blue trace) and PCSNs as used in panels (b,c)
(red trace). For the continuous network $N=50$, for the PCSNs $\gamma_{s}=0.01,N=70,J=50$.
The PCSN has a mean spike rate of $441$ Hz

\subsubsection*{Figure G}

Fig. Ga: The panel shows the last $100$s of the PCSN simulation displayed
in Fig.~Fd. Total duration of the recall phase is $10000$s, Fig.~Fd
shows the first $100$s.

Fig. Gb: We use the same network parameters as in Fig.~Fd and Fig.~Ga,
displayed are $10$ learning trials with different randomly chosen
initial connectivity. We compute a phase-shift corrected version of
the RMS error in a sliding window of size $100$s, for different starting
points $\tau$ of the sliding window. $\tau$ is in the range from
$1$s to $10000$s with step size $1$s. The phase-shift corrected
version of the RMS error is computed by $\sqrt{\frac{1}{100\text{s}}\intop_{\tau}^{\tau+100\text{s}}\left(\text{signal}(\widetilde{t})-\text{target}(\widetilde{t}+\Delta)\right)^{2}d\widetilde{t}}$
with the phase-shift $\Delta$ of the target signal chosen such that
the integral is minimal. 

Fig. Gc: The parameters are the same as in Fig.~3h.
\renewcommand\refname{Supplemental references}

\end{document}